\documentclass[12pt,letterpaper]{article}
\usepackage[margin=0.9in]{geometry}
\usepackage{amsmath,amssymb,amsthm,mathtools,bm}
\allowdisplaybreaks
\usepackage[T1]{fontenc}
\usepackage{newtxtext}
\usepackage{newunicodechar}
\newunicodechar{α}{\ensuremath{\alpha}}
\newunicodechar{β}{\ensuremath{\beta}}
\newunicodechar{γ}{\ensuremath{\gamma}}
\newunicodechar{δ}{\ensuremath{\delta}}
\newunicodechar{ε}{\ensuremath{\varepsilon}}
\newunicodechar{κ}{\ensuremath{\kappa}}
\newunicodechar{λ}{\ensuremath{\lambda}}
\newunicodechar{μ}{\ensuremath{\mu}}
\newunicodechar{σ}{\ensuremath{\sigma}}
\newunicodechar{τ}{\ensuremath{\tau}}
\newunicodechar{Δ}{\ensuremath{\Delta}}
\newunicodechar{Λ}{\ensuremath{\Lambda}}
\newunicodechar{Ω}{\ensuremath{\Omega}}
\newunicodechar{Σ}{\ensuremath{\Sigma}}
\newunicodechar{χ}{\ensuremath{\chi}}
\newunicodechar{ }{\,}
\usepackage{setspace}
\usepackage{graphicx}
\usepackage{booktabs}
\usepackage{longtable}
\usepackage{array}
\usepackage{multirow}
\usepackage{caption}
\usepackage{pdflscape}
\usepackage{threeparttable}
\usepackage[colorlinks=true,linkcolor=blue!50!black,urlcolor=blue!50!black,citecolor=blue!50!black]{hyperref}
\usepackage{enumitem}
\setlist{nosep}
\usepackage{xcolor}
\usepackage{titling}
\usepackage{fancyhdr}
\theoremstyle{definition}

\theoremstyle{remark}

\title{Partial Identification with Multiple Nonlinear Measurements of a Latent Regressor}
\author{Burhan Ogut\thanks{American Institutes for Research. Corresponding author: bogut@air.org.}
\and Michelle Yin\thanks{Northwestern University, School of Education and Social Policy.}}
\date{\today}

\begin{document}
\maketitle

\begin{abstract}
We study linear regression when the regressor is latent and observed only through multiple noisy measurements, each a smooth but possibly nonlinear function of the latent variable. The problem arises acutely in the measurement of occupational exposure to artificial intelligence, where competing scores yield downstream estimates that differ by a factor of eleven. A regression on any single measurement recovers a source-specific coefficient rather than the structural one. With several measurements, the nonlinearity itself becomes useful. We fix the latent scale by requiring the consensus measurement function to be linear and bound the remaining curvature heterogeneity across sources relative to slope. Under this bound, the structural coefficient lies in a closed-form interval centered at a symmetric cross-source estimator. The interval is invariant to unknown source loadings, and its half-width is second order in the curvature bound and sharp to the same order. With at least four measurements, the bound itself is estimable from the joint distribution of the sources through a split-instrument auxiliary regression, and confidence intervals based on the Imbens and Manski procedure with the Stoye critical value attain uniform coverage over the curvature class, including at the point-identified boundary. The application matches six exposure measures to an American Community Survey panel of 8.88 million person-year observations for 2015 to 2024. The post-2022 employment coefficient changes sign between the language-model measures and the Webb patent-text measure, and an ex ante factor-analytic rule separates the Webb measure as a distinct construct. The five retained sources yield a loading-invariant consensus coefficient of -0.239, with a partial-identification half-width of 1.23 percent of the point estimate, or 1.88 percent at the one-sided 95 percent upper bound on the curvature. We read the application as measurement reconciliation rather than as a causal estimate of AI displacement.

\end{abstract}

\thispagestyle{fancy}
\newpage

\section{Introduction}
The empirical literature on artificial intelligence and labor-market outcomes typically assigns each occupation an exposure measure and relates that measure to worker outcomes in panel, difference-in-differences, or shift-share regressions. Several distinct families of occupational exposure measures have become available. Large-language-model measures score O*NET task descriptions according to rubrics designed to capture the applicability of current AI systems (Eloundou et al., 2024). Felten, Raj, and Seamans (2021) aggregate occupational abilities using a profile of progress in AI capabilities. Webb (2020) measures textual overlap between occupational task descriptions and AI-related patents. More recent measures use platform-level conversation shares from Anthropic, OpenAI, and Microsoft systems to infer occupational exposure (Handa et al., 2025; Chatterji et al., 2025; Tomlinson et al., 2025). Composite indices combine several of these inputs (Massenkoff and McCrory, 2026).

These measures are often treated as alternative observations of the same underlying occupational exposure. In practice, however, they can imply substantially different downstream estimates when merged into the same labor-market data. In an American Community Survey panel with approximately 8.88 million workers observed from 2015 to 2024, the estimated post-2022 employment coefficient differs by a factor of eleven across the six single-source exposure measures considered below. The coefficient also changes sign between the language-model-based measures and the patent-text measure of Webb (2020). The choice of exposure measure is therefore not only a normalization or labeling decision. It is a source of specification uncertainty that directly affects the reported labor-market coefficient. Conventional standard errors from a regression using any one exposure measure do not account for this uncertainty.

This paper studies the econometric problem that gives rise to this disagreement. We model the available exposure measures as noisy nonlinear measurements of a common latent regressor, with each source applying its own smooth monotone transformation to the latent variable. A regression on any single source therefore estimates a source-specific object rather than the structural coefficient of interest, and the discrepancy between the two is governed by the slope and curvature of that source's measurement function. Without restrictions on the measurement equation, one source is not sufficient for point identification.

The main result shows that multiple measurements can be used to obtain informative partial identification. We consider a class of measurement functions whose curvature, measured relative to slope, is uniformly bounded. Under this restriction, the structural coefficient is confined to an interval with closed-form endpoints, centered at a symmetric cross-source minimum-distance estimator. The endpoints do not depend on the unknown source-specific loadings, and the half-width is second order in the curvature bound for any latent distribution satisfying the stated moment conditions. This second-order property follows from the symmetry of the construction, which removes the first-order skewness term that appears in asymmetric instrumental-variable estimators. The interval is sharp to the same order.

The curvature bound is not treated as a fixed tuning parameter. Because the latent regressor is observed only through the sources, its scale is fixed by a normalization under which the consensus measurement function is linear, and the bound restricts source-specific departures from that scale. With at least four measurements, the bound is then estimable from the joint distribution of the sources. We use a split-instrument auxiliary regression. One leave-out average and its square enter as regressors, while a disjoint leave-out average and its square serve as instruments. This construction corrects the errors-in-variables attenuation that would affect a direct auxiliary regression. The resulting estimated endpoints are jointly asymptotically normal. We construct confidence intervals for the partially identified coefficient using the Imbens and Manski procedure with the Stoye critical value. The coverage result is uniform over the curvature class, including at the boundary where the model becomes point identified.

The closest antecedents are Armstrong and Kolesár (2018, 2020), who develop optimal honest inference when a regression function has a second derivative bounded by a known constant. The restriction used here is analogous, but it is imposed on measurement functions of a latent regressor rather than on a conditional mean. The main difference is that the multi-source structure makes the relevant curvature bound estimable from the data. The paper also relates to the partial-identification and moment-inequality literatures, including Manski (2003), Tamer (2010), Chen, Christensen, and Tamer (2018), Kaido, Molinari, and Stoye (2019), Canay and Shaikh (2017), and Molinari (2020), and to the deconvolution literature, including Hu and Schennach (2008) and Schennach (2016).

The framework differs from discrete-class correction models such as Dawid-Skene maximum likelihood and from related multi-rater estimators, including the obviously-related-IV estimator of Gillen, Snowberg, and Yariv (2019). Those approaches typically require conditional independence of rater errors at the distributional level. The present approach requires only second-moment orthogonality across sources. In the empirical application, the two families of corrections yield similar downstream coefficients.

The empirical analysis applies the method to six measures of occupational AI exposure matched to the American Community Survey from 2015 to 2024. An ex ante factor-analytic rule retains the four language-model measures and the Felten et al. index, but excludes the Webb patent-text measure, which behaves as a distinct construct in the data. For the five retained measures, the proposed loading-invariant estimator yields a consensus coefficient with a narrow partial-identification band. The half-width is 1.23 percent of the point estimate at the estimated curvature bound and 1.88 percent when the curvature bound is replaced by its one-sided 95 percent upper confidence bound. The remaining curvature uncertainty is therefore small relative to sampling uncertainty.

The empirical exercise is intended as a measurement-reconciliation application rather than as a causal estimate of AI displacement. Standard parallel-trends diagnostics reject for the post-2022 design, with the rejection concentrated in the COVID-era composition reversal in 2020 and 2021. We therefore report COVID-excluded specifications alongside the main estimates and interpret the exercise as evidence on how multiple imperfect exposure measures can be combined and disciplined.

The methodological contribution is a closed-form, data-driven partial-identification interval for a structural coefficient when the regressor is latent and observed through multiple noisy nonlinear measurements. The interval is centered at a symmetric minimum-distance estimator, is invariant to unknown source loadings, and has width governed by an estimable bound on relative curvature. The empirical contribution is a corrected occupation-level AI-exposure score for 682 six-digit Standard Occupational Classification codes, together with partial-identification bands. We also provide a LASSO-sparse O*NET-feature own-measure that approximates the corrected score without using any large-language-model input. Both objects are released as companion files for use in downstream empirical work.

Although the application concerns AI exposure, the identification problem is more general. The same structure arises whenever a latent treatment, exposure, or rating variable is observed through several noisy sources, each of which may apply a different smooth monotone transformation. Examples include ESG ratings, which differ substantially across providers and can lead to provider-dependent estimates in firm-value or cost-of-capital regressions, and LLM-based annotation tasks that produce multiple sentiment scores, simulated survey responses, or qualitative codings. The results below apply when at least four noisy measurements are available and the analyst can estimate the curvature bound using the proposed split-instrument diagnostic.

The remainder of the paper is organized as follows. Section 2 introduces the multi-source measurement model and states the identifying assumptions. Section 3 establishes the partial-identification result. Section 4 defines the estimator and the curvature diagnostic. Section 5 develops the asymptotic theory and the uniformly valid Imbens-Manski confidence interval. Section 6 reports Monte Carlo evidence. Section 7 applies the method to the occupational AI-exposure data. Section 8 describes the released score table and the LASSO-sparse own-measure. Section 9 concludes.

\section{The Multi-Source Measurement Model}
\subsection{Setup}
Units are indexed by \(i = 1,\ldots,n\). Each unit is assigned to an occupation \(o(i) \in \{ 1,\ldots,O\}\), defined at the six-digit Standard Occupational Classification level. Let \(E_{i}\) denote the latent scalar exposure of unit \(i\). The structural equation is

\begin{equation}
Y_i = \beta E_i + X_i'\theta + \varepsilon_i, \qquad \mathbb{E}[\varepsilon_i \mid E_i, X_i] = 0, \label{eq:struct}
\end{equation}

where \(Y_{i}\) is the outcome, \(X_{i}\) is a vector of controls, and \(\beta\) is the parameter of interest.

The latent regressor is not observed directly. Instead, the analyst observes \(S \geq 4\) noisy measurements,

\begin{equation}
M_i^{(s)} = \mu_s(E_i) + U_i^{(s)}, \qquad s = 1, \ldots, S, \label{eq:meas}
\end{equation}

where \(\mu_{s}\) is a source-specific measurement function and \(U_{i}^{(s)}\) is source-specific measurement error. The functions \(\mu_{s}\) are allowed to differ across sources. They are assumed to be smooth and monotone, but are otherwise unrestricted. This formulation allows each observed measure to be a nonlinear transformation of the same latent exposure.

In the empirical application, the sources are the exposure measures that pass the construct screen described in Section 7. The extension in which sources also contain platform-specific selection components is developed in Appendix B. Throughout the main text, variables are understood as residualized with respect to \(X_{i}\), following the Frisch-Waugh-Lovell theorem. This convention permits the exposition to focus on the scalar latent regressor while retaining the original control structure.

The model includes the linear measurement case as a limiting case. When the curvature of each \(\mu_{s}\) is zero, the sources differ only in location, scale, and measurement noise. In that case, the multi-source problem reduces to a standard linear latent-regressor setting. The contribution of the paper is to characterize what remains identified when the measurement functions are allowed to be nonlinear but their relative curvature is bounded.

A discrete special case connects the model to the multi-rater misclassification literature. Suppose that a rater discretizes the latent variable into the ordinal categories used in the Eloundou et al. (2024) rubric and reports a binary collapse of those categories. The corresponding measurement function is then approximately a threshold rule, with curvature concentrated around a rater-specific cutoff. The attenuation parameter used in latent-class models can be interpreted as the discrete limit of the smooth measurement framework developed here. Section 7 uses this connection to compare the proposed estimator with Dawid-Skene, ORIV, and Imbens-Manski corrections under a discrete-rater specialization.

\subsection{Assumptions}
Identification is based on the following assumptions.

\textbf{Assumption 1. Latent normalization.}\\
The latent variable satisfies

\[\mathbb{E}[ E_{i}] = 0,\operatorname{Var}(E_{i}) = 1.\]

Let

\[\kappa_{3} = \mathbb{E}[ E_{i}^{3}],\kappa_{4} = \mathbb{E}[ E_{i}^{4}].\]

\textbf{Assumption 2. Measurement error.}\\
For each source \(s\),

\[\begin{aligned}
\mathbb{E}\left[ U_{i}^{(s)} \mid E_{i},X_{i},\{ U_{i}^{(t)}:t \neq s\} \right] = 0 \\ \mathbb{E}\left[ U_{i}^{(s)} \mid \varepsilon_{i},E_{i},X_{i} \right] = 0
\end{aligned}\]

and

\[\mathbb{E}\left[ (U_{i}^{(s)})^{2} \mid E_{i} \right] = \sigma_{s}^{2} < \infty.\]

\textbf{Assumption 3. Smooth monotone measurement.}\\
Each \(\mu_{s}\) is three times continuously differentiable on the support of \(E_{i}\). Moreover,

\[\mu_{s}'(0) = c_{s} > 0.\]

Around the normalization point,

\begin{equation}
\mu_s(E) = a_s + c_s E + \tfrac{1}{2}\gamma_s E^2 + R_s(E), \label{eq:mu}
\end{equation}

where \(R_{s}(E)\) denotes the third-order Taylor remainder.

\textbf{Assumption 4. Relative curvature.}\\
There exists a finite constant \(B\) such that, for every source \(s\),

\[\lvert \gamma_{s} \rvert \leq Bc_{s}.\]

In addition, there exists a fixed finite constant \(C\) such that

\[\underset{E}{\sup} \lvert \mu_{s}^{'''}(E) \rvert \leq CB^{2}c_{s}\]

for every \(s\).

\textbf{Assumption 5. Sampling and moments.}\\
The observations are independent and identically distributed. The latent variable satisfies

\[\mathbb{E}[ E_{i}^{8}] < \infty.\]

The measurement errors have finite fourth moments. The matrix \(\mathbb{E}[ X_{i}X_{i}']\) is nonsingular. The cluster-dependent extension is given in Section 5.

\textbf{Assumption 6. Regularity of the estimated curvature bound.}\\
The source attaining

\[\underset{1 \leq s \leq S}{\max} \lvert \gamma_{s}/c_{s} \rvert\]

is unique.

\subsection{Discussion of the assumptions}
Assumption 1 is a normalization. Since the latent variable is not observed, its location and scale cannot be separately identified from the intercepts and loadings of the measurement functions. The normalization fixes this indeterminacy and defines the scale on which \(\beta\) is interpreted.

Assumption 2 imposes mean independence restrictions on the measurement errors. Part (a) rules out systematic dependence between one source\textquotesingle s measurement error and the latent variable or the other sources\textquotesingle{} errors. Part (b) rules out dependence between source-level measurement errors and the structural disturbance. Part (c) imposes homoskedasticity of the measurement error with respect to the latent variable. The homoskedasticity restriction is used in the auxiliary curvature regression. It can be relaxed to bounded conditional variance at the cost of more notation and a longer proof.

The restriction in Assumption 2 is weaker than the full conditional independence conditions often used in discrete latent-class models. Those models typically require independence of rater errors at the distributional level. Here the identifying arguments use conditional mean restrictions and second moments.

Assumption 3 allows the loadings \(c_{s}\) to differ across sources. This heterogeneity is essential. Standardizing the observed measurements does not generally make their latent loadings equal, because the observed variances combine signal variation, nonlinear transformation, and source-specific noise. The estimator introduced below is constructed so that these unknown loadings cancel from the identifying ratio.

The latent regressor is observed only through the sources, so its scale requires a normalization beyond Assumption 1. A transformation of \(E\) that is common to every measurement function cannot be distinguished from a redefinition of the latent variable: the joint distribution of the sources is unchanged when each \(\mu_{s}\) is composed with the same smooth monotone map and \(E\) is relabeled accordingly. We therefore normalize the latent scale so that the equally weighted average of the measurement functions is linear in \(E\), which sets \(\sum_{s}\gamma_{s} = 0\) in the expansion of Assumption 3. Under this normalization the curvature parameters \(\gamma_{s}\) measure source-specific departures from the consensus scale, and \(\beta\) is the coefficient on the exposure expressed in that scale. The normalization matters for interpretation. If the outcome equation is instead taken to be linear in some other latent scale, the common curvature component separating the two scales is not identifiable from the sources and is not covered by the interval derived below. Section 4.2 shows that the split-instrument diagnostic estimates precisely the departures that the normalization leaves free.

Assumption 4 bounds curvature relative to slope. This scale-free form is important because the units of each observed source are arbitrary. A source with a larger loading is allowed to have proportionally larger second derivative. The additional restriction on the third derivative ensures that the Taylor remainder is of smaller order relative to the leading curvature term. This condition is satisfied, for example, by a class of transformations of the form

\[\mu_{s}(E) = c_{s}\frac{\phi(h_{s}E)}{h_{s}}, \lvert h_{s} \rvert \leq B,\]

for a fixed smooth function \(\phi\) with bounded derivatives. In that case, the second derivative is of order \(c_{s}B\), while the third derivative is of order \(c_{s}B^{2}\). The condition therefore ties the orders of the second and third derivatives to a common curvature scale rather than treating them as unrelated primitive constants.

Assumption 6 is used only because the estimated curvature bound is defined as a maximum over sources. The uniqueness condition is generic. If the maximum is attained by more than one source, the asymptotic distribution can be obtained using directional differentiability arguments, as in Fang and Santos (2019).

\subsection{Half-averages and source splitting}
The identification argument uses averages over disjoint subsets of sources. Partition the source set into two disjoint subsets \(S_{1}\) and \(S_{2}\), each of size \(K\) when \(S\) is even. When \(S\) is odd, the two subsets are chosen to have approximately equal sizes. Define

\[{\bar{M}}_{i}^{(j)} = \frac{1}{K}\sum_{s \in S_{j}}^{}M_{i}^{(s)},j = 1,2.\]

Using the expansion in Assumption 3,

\begin{equation}
\bar M_i^{(j)} = \bar a_j + \bar c_j E_i + \tfrac{1}{2}\bar\gamma_j E_i^2 + \bar U_i^{(j)} + \bar R_i^{(j)}, \qquad j = 1, 2, \label{eq:subavg}
\end{equation}

where \({\bar{a}}_{j}\), \({\bar{c}}_{j}\), and \({\bar{\gamma}}_{j}\) are the corresponding averages of \(a_{s}\), \(c_{s}\), and \(\gamma_{s}\) over subset \(S_{j}\), and \({\bar{U}}_{i}^{(j)}\) and \({\bar{R}}_{i}^{(j)}\) are the corresponding averages of measurement errors and Taylor remainders.

Let

\[h_{j} = \frac{{\bar{\gamma}}_{j}}{{\bar{c}}_{j}}\]

denote the relative curvature of subset \(j\). By Assumption 4 and \(c_{s} > 0\),

\[\lvert h_{j} \rvert \leq B.\]

Assumption 2 implies that the measurement errors in the two subset averages are conditionally mean independent of the latent variable and of each other. This half-subset representation is the basic object used in the identification and estimation results below.

\section{Identification}
This section derives the identified set for the structural coefficient. Section 3.1 gives the population limits of two estimators under the quadratic approximation implied by Assumption 4. The first is an asymmetric two-stage least-squares ratio based on one measurement average as an instrument for the other. The second is a symmetric cross-source minimum-distance estimator. Section 3.2 uses these population limits to construct a closed-form partial-identification interval for \(\beta\). Section 3.3 records the Gaussian benchmark. Section 3.4 relates the result to existing partial-identification and measurement-error frameworks.

\subsection{Population limits}
By Assumptions 3 and 4, each measurement function admits the expansion

\begin{equation}
\mu_s(E) = a_s + c_s E + \tfrac{1}{2}\gamma_s E^2 + R_s(E),
\end{equation}

where \(\lvert \gamma_{s} \rvert \leq Bc_{s}\) and the Taylor remainder satisfies

\[\lvert R_{s}(E) \rvert \leq \frac{1}{6}CB^{2}c_{s} \lvert E \rvert^{3}.\]

Using the source-splitting construction in Section 2.4, the average measurement in subset \(j \in \{ 1,2\}\) can be written as

\begin{equation}
\bar M_i^{(j)} = \bar a_j + \bar c_j E_i + \tfrac{1}{2}\bar\gamma_j E_i^2 + \bar U_i^{(j)} + \bar R_i^{(j)}, \qquad j = 1, 2,
\end{equation}

where \({\bar{c}}_{j}\) and \({\bar{\gamma}}_{j}\) are the corresponding subset averages of \(c_{s}\) and \(\gamma_{s}\). Let

\[h_{j} = \frac{{\bar{\gamma}}_{j}}{{\bar{c}}_{j}}\]

denote the relative curvature of subset \(j\). By Assumption 4, \(\lvert h_{j} \rvert \leq B\).

Define the population covariances

\[\sigma_{Yj} = \operatorname{Cov}(Y_{i},{\bar{M}}_{i}^{(j)}), \qquad \sigma_{12} = \operatorname{Cov}({\bar{M}}_{i}^{(1)},{\bar{M}}_{i}^{(2)}).\]

Two ratios are useful. The asymmetric two-stage least-squares estimand is

\begin{equation}
\beta_{2SLS} = \frac{\sigma_{Y2}}{\sigma_{12}},
\end{equation}

where \({\bar{M}}_{i}^{(2)}\) is used as an instrument for \({\bar{M}}_{i}^{(1)}\). The symmetric estimator is defined by

\begin{equation}
\beta_{sym} = \operatorname{sign}\!\left(\frac{\sigma_{Y2}}{\sigma_{12}}\right)\left(\frac{\sigma_{Y1}\,\sigma_{Y2}}{\sigma_{12}}\right)^{1/2},
\end{equation}

whenever the radicand is positive. This condition holds for small enough \(B\) when \(\beta \neq 0\) and \(\sigma_{12} > 0\).

\textbf{Proposition 1. Population limits.}\\
Suppose Assumptions 1 through 5 hold. Let

\[D = 1 + \frac{1}{2}(h_{1} + h_{2})\kappa_{3} + \frac{1}{4}h_{1}h_{2}(\kappa_{4} - 1),\]

and assume \(D > 0\). Then

\begin{equation}
\operatorname{plim}_{n\to\infty}\beta_{2SLS} = \frac{\beta}{\bar c_1}\,\frac{1 + \tfrac{1}{2}h_2\kappa_3}{D} + O(B^2),
\end{equation}

and

\begin{equation}
\left(\operatorname{plim}_{n\to\infty}\beta_{sym}\right)^2 = \beta^2\,\frac{1 + \tfrac{1}{2}(h_1+h_2)\kappa_3 + \tfrac{1}{4}h_1 h_2\kappa_3^2}{D} + O(B^3),
\end{equation}

Proposition 1 shows the role of symmetry. The asymmetric ratio depends on the loading of the endogenous measurement, \({\bar{c}}_{1}\), and contains a first-order term in the skewness of the latent regressor. The symmetric estimator eliminates the unknown loadings because the loadings enter the three covariances multiplicatively and cancel in the ratio. Its leading deviation from \(\beta\) is second order in the relative curvatures. The cancellation occurs because the first-order skewness terms enter the numerator product \(\sigma_{Y1}\sigma_{Y2}\) and denominator \(\sigma_{12}\) in the same way.

\subsection{The identified set}
The relative curvatures \(h_{1}\) and \(h_{2}\) are not separately identified. Identification instead uses the common bound \(\lvert h_{j} \rvert \leq B\) implied by Assumption 4. The following theorem converts the population expansion in Proposition 1 into a closed-form partial-identification interval.

\textbf{Theorem 1. Curvature-bounded partial identification.}\\
Suppose Assumptions 1 through 5 hold. Then

\[\lvert \beta - \operatorname{plim}_{n \rightarrow \infty}\beta_{sym} \rvert \leq \lvert \operatorname{plim}_{n \rightarrow \infty}\beta_{sym} \rvert \Delta + O(B^{3}),\]

where

\begin{equation}
\Delta = \tfrac{1}{8}B^2(\kappa_4 - 1 - \kappa_3^2). \label{eq:delta}
\end{equation}

Moreover,

\[\kappa_{4} - 1 - \kappa_{3}^{2} \geq 0.\]

This term is the variance of the residual from the linear projection of \(E_{i}^{2}\) on \(\left( 1,E_{i} \right)\). The bound is sharp to second order in \(B\): for each value in the interval \([ - \Delta,\Delta]\), there exists an admissible pair \((h_{1},h_{2}) \in [ - B,B]^{2}\) that attains the corresponding second-order deviation, up to an \(O(B^{3})\) remainder.

Theorem 1 gives the identified set

\begin{equation}
\mathcal{B}(B) = \left[\,\frac{\beta_{sym}}{1+\Delta},\ \frac{\beta_{sym}}{1-\Delta}\,\right] + O(B^3), \label{eq:idset}
\end{equation}

with the endpoints interpreted after replacing the population covariances by their population limits. Equivalently, for small \(\Delta\),

\begin{equation}
\mathcal{B}(B) = \left[\,\beta_{sym}(1-\Delta),\ \beta_{sym}(1+\Delta)\,\right] + O(B^3),
\end{equation}

The interval has four properties that are important for the empirical implementation. First, its width is second order in the curvature bound. Second, it is invariant to the unknown source-specific loadings. Third, it reduces continuously to point identification as \(B \downarrow 0\). Fourth, under the normalization \(\sum_{s}\gamma_{s} = 0\) and an equal split, the subset curvatures satisfy \({\bar{c}}_{1}h_{1} = -{\bar{c}}_{2}h_{2}\), so the product \(h_{1}h_{2}\) is nonpositive and the second-order deviation has a determinate sign; the interval does not impose this restriction and is conservative in that respect. Imposing it would remove the half of the interval on the side of larger magnitude, at the cost of tying the validity of the band to the normalization and to the balance of the split. Because the reported half-width is already an order of magnitude below the sampling uncertainty, we retain the two-sided interval.

For comparison, the asymmetric estimator yields a wider bound under the additional restriction of homogeneous loadings.

\textbf{Corollary 1. Asymmetric estimator under homogeneous loadings.}\\
Suppose Assumptions 1 through 5 hold and \(c_{s} = c\) for all \(s\). Then

\[\lvert \beta - c\text{ }\operatorname{plim}_{n \rightarrow \infty}\beta_{2SLS} \rvert \leq \lvert c\text{ }\operatorname{plim}_{n \rightarrow \infty}\beta_{2SLS} \rvert \Delta_{a} + O(B^{2}),\]

where

\begin{equation}
\Delta_a = \tfrac{1}{2}B\,|\kappa_3| + \tfrac{1}{4}B^2\,|\kappa_4 - 1|.
\end{equation}

Thus, the asymmetric construction requires an additional loading restriction and has a first-order component whenever the latent distribution is skewed. We report it in the empirical application as a robustness specification rather than as the baseline estimator.

\subsection{Gaussian benchmark}
The next corollary gives a simple benchmark for the width of the identified set.

\textbf{Corollary 2. Gaussian latent exposure.}\\
Suppose Assumptions 1 through 5 hold and \(E_{i} \sim N(0,1)\). Then \(\kappa_{3} = 0\) and \(\kappa_{4} = 3\), so that

\[\Delta = \frac{B^{2}}{4} + O(B^{4}).\]

Under Gaussian latent exposure, the symmetric half-width is quadratic in \(B\). The corresponding asymmetric half-width in Corollary 1 is \(B^{2}/2\) when \(\kappa_{3} = 0\). The Gaussian case is therefore useful as a benchmark, although the results above do not require normality.

\subsection{Relation to existing identification frameworks}
Theorem 1 is closest to the bounded-curvature approach of Armstrong and Kolesár (2018, 2020), who study honest inference when a regression function has a bounded second derivative. The restriction here is analogous but applies to measurement functions of a latent regressor rather than to a conditional mean. The curvature bound is also expressed relative to the source-specific loading, which makes it invariant to the scale of each observed measurement. A further difference is that the multi-source structure permits the bound to be estimated from the joint distribution of the measurements, rather than fixed by the analyst.

The result also relates to the partial-identification literature following Manski (2003). In that literature, bounds often enter as external restrictions on the magnitude of an unobserved bias or counterfactual object. Here the bound is tied to the curvature of the measurement functions and can be disciplined by the auxiliary regression in Section 4. Relative to the moment-inequality and set-inference literature, including Tamer (2010), Canay and Shaikh (2017), Chen, Christensen, and Tamer (2018), Kaido, Molinari, and Stoye (2019), and Molinari (2020), the present setting yields closed-form endpoints. Inference therefore reduces to inference on an estimated interval.

The framework differs from deconvolution approaches such as Hu and Schennach (2008) and Schennach (2016), which obtain identification by restricting the distribution of the latent variable and the measurement errors. The present approach instead uses multiple measurements and a curvature restriction, without requiring full knowledge of the error distribution. The number of sources matters: with one, the curvature bound would have to be supplied externally; with two or three, additional structure is needed to estimate the relevant curvature object; the full construction developed here requires at least four.

The discrete-rater model discussed in Section 2 is a limiting case. In that setting, a rater reports a discretized version of the latent variable, and the measurement function approximates a step function with rater-specific thresholds. Latent-class models, including Dawid and Skene (1979), identify class probabilities and rater-specific confusion matrices under conditional independence of rater errors. Related multi-rater estimators, such as the obviously-related-IV estimator of Gillen, Snowberg, and Yariv (2019), also exploit cross-rater variation. These approaches impose conditional independence at the distributional level. The present framework requires only the mean and second-moment restrictions in Assumption 2. It therefore covers a different part of the same general measurement problem: continuous latent variables observed through several noisy, possibly nonlinear sources.

\subsection{Failure modes and scope of the result}
Theorem 1 is a local curvature result. Its accuracy depends on the second-order approximation capturing the leading distortion in each measurement function. If \(B\) is large, the \(O(B^{3})\) remainder may no longer be negligible, and the closed-form interval should be interpreted as a second-order approximation rather than a sharp finite-\(B\) characterization.

The result also relies on the cross-source orthogonality restrictions in Assumption 2. Correlated source errors can affect the covariance \(\sigma_{12}\) and thereby bias the symmetric estimator. The relevant empirical diagnostic is the residual correlation matrix across sources, reported in Section 7.

Finally, the requirement \(S \geq 4\) is substantive. The split-instrument curvature diagnostic in Section 4 requires disjoint source subsets, with enough measurements to form leave-out averages that are not mechanically identical. When fewer measurements are available, the curvature bound must be treated as an external sensitivity parameter or supplemented by additional identifying restrictions.

The sharpness statement in Theorem 1 is conditional on the information summarized by \(\left( B,\kappa_{3},\kappa_{4} \right)\). In this sense, the interval is sharp to second order for the information set used by the analyst. Additional restrictions on the measurement functions or the distribution of the latent variable could further reduce the identified set, but those restrictions are not imposed here.

\section{Estimation}
Section 3 characterizes the identified set for the structural coefficient in terms of the symmetric cross-source estimand and the curvature bound. This section describes how these objects are estimated. Section 4.1 defines the sample analogue of the symmetric estimator. Section 4.2 describes the split-instrument auxiliary regression used to estimate the relative-curvature bound. Section 4.3 explains the estimation of the standardized moments entering the half-width. Section 4.4 discusses partition choice, and Section 4.5 reports the diagnostic role of first-stage strength and residual dependence.

\subsection{The symmetric estimator}
Let \(M_{i}^{(s)}\) denote source \(s\)\textquotesingle s measurement of the latent regressor. For a partition of the source set into two disjoint subsets \(S_{1}\) and \(S_{2}\), define the subset averages

\[{\bar{M}}_{i}^{(j)} = \frac{1}{\lvert S_{j} \rvert}\sum_{s \in S_{j}}^{}M_{i}^{(s)},j = 1,2.\]

The estimator uses the three sample covariances corresponding to the population quantities in Section 3. All variables are first residualized with respect to the controls \(X_{i}\). Let \({\widetilde{Y}}_{i}\) and \({\widetilde{\bar{M}}}_{i}^{(j)}\) denote the resulting Frisch-Waugh-Lovell residuals. Define

\[{\widehat{\sigma}}_{Yj} = \frac{1}{n}\sum_{i = 1}^{n}{\widetilde{Y}}_{i}{\widetilde{\bar{M}}}_{i}^{(j)},j = 1,2,\]

and

\[{\widehat{\sigma}}_{12} = \frac{1}{n}\sum_{i = 1}^{n}{\widetilde{\bar{M}}}_{i}^{(1)}{\widetilde{\bar{M}}}_{i}^{(2)}.\]

The symmetric estimator is

\begin{equation}
\hat\beta_{sym} = \operatorname{sign}\!\left(\frac{\hat\sigma_{Y2}}{\hat\sigma_{12}}\right)\left(\frac{\hat\sigma_{Y1}\,\hat\sigma_{Y2}}{\hat\sigma_{12}}\right)^{1/2}, \label{eq:symest}
\end{equation}

whenever the radicand is positive. Under the conditions in Section 3, this event has probability approaching one when \(\beta \neq 0\) and the curvature bound is sufficiently small.

In the empirical application, the observed measurements are occupation-level exposure scores. We first aggregate source readings to the occupation level and standardize each source on the analysis sample. This standardization places the measurements on a common numerical scale for implementation. It does not impose equal latent loadings, since the symmetric estimator is invariant to the unknown source-specific loadings.

For the difference-in-differences specification used in the application, the measured regressor is the interaction between the occupation-level exposure proxy and the post period. Thus, \({\bar{M}}_{i}^{(j)}\) is replaced by

\[{\bar{M}}_{o(i)}^{(j)} \cdot Post_{t(i)}.\]

The residualization step absorbs the controls and fixed effects used in the downstream regression. In the application these include occupation, state, and year fixed effects. The same covariance formula then defines the estimator.

The sampling variance is obtained by applying the delta method to the vector

\[\left( {\widehat{\sigma}}_{Y1},{\widehat{\sigma}}_{Y2},{\widehat{\sigma}}_{12} \right).\]

When observations are clustered, the covariance matrix of these sample moments is estimated using cluster-robust methods. In the empirical application, clustering is at the six-digit SOC level.

For comparison, we also report the asymmetric two-stage least-squares estimator

\[{\widehat{\beta}}_{2SLS} = \frac{{\widehat{\sigma}}_{Y2}}{{\widehat{\sigma}}_{12}},\]

which corresponds to using one subset average as an instrument for the other. As shown in Corollary 1, this estimator is not loading-invariant without additional restrictions. It is therefore used as a robustness specification rather than as the baseline estimator.

\subsection{Estimating the curvature bound}
The half-width of the identified set depends on the relative-curvature bound \(B\) and on the standardized moments of the latent regressor. The bound \(B\) is estimated from an auxiliary regression that exploits the availability of multiple measurements.

A direct auxiliary regression of source \(s\) on a leave-out average and its square is attenuated because the leave-out average is itself measured with error. This attenuation biases the estimated curvature toward zero and would lead to an interval that is too narrow. To avoid this problem, we use a split-instrument construction.

Fix a source \(s\). Partition the remaining sources into two disjoint subsets \(A_{s}\) and \(B_{s}\), chosen to have approximately equal size. Let

\[Z_{iA}^{(s)} = \frac{1}{\lvert A_{s} \rvert}\sum_{r \in A_{s}}^{}M_{i}^{(r)}\]

and

\[Z_{iB}^{(s)} = \frac{1}{\lvert B_{s} \rvert}\sum_{r \in B_{s}}^{}M_{i}^{(r)}.\]

For each source \(s\), estimate the auxiliary equation

\begin{equation}
M_i^{(s)} = b_{0s} + b_{1s} Z_{iA}^{(s)} + b_{2s}\left(Z_{iA}^{(s)}\right)^2 + e_i^{(s)}, \label{eq:aux}
\end{equation}

by two-stage least squares, using

\[1,Z_{iB}^{(s)},\left( Z_{iB}^{(s)} \right)^{2}\]

as instruments. Under Assumption 2, the measurement error in \(Z_{iB}^{(s)}\) is orthogonal to the measurement error in \(Z_{iA}^{(s)}\). The instrumental-variables regression therefore removes the errors-in-variables attenuation in the quadratic coefficient.

The source-specific relative-curvature estimator is

\begin{equation}
\hat g_s = 2\hat b_{2s}\,\frac{\sqrt{\operatorname{Cov}\!\left(Z_{iA}^{(s)}, Z_{iB}^{(s)}\right)}}{\hat b_{1s}}, \label{eq:curv}
\end{equation}

The estimated curvature bound is

\begin{equation}
\widehat B = \max_{1 \leq s \leq S}|\hat g_s|.
\end{equation}

The factor \(2{\widehat{b}}_{2s}\) converts the coefficient on the squared proxy into a second-derivative scale. The square-root covariance term estimates the leading latent loading of the leave-out proxy under the balanced split. Division by \({\widehat{b}}_{1s}\) then expresses curvature relative to the source-specific slope. The population limit of \({\widehat{g}}_{s}\) is \(\left( \gamma_{s}/c_{s} - {\bar{\gamma}}_{A}/{\bar{c}}_{A} \right)\left( {\bar{c}}_{B}/{\bar{c}}_{A} \right)^{1/2}\) up to terms of order \(B^{2}\), where \({\bar{\gamma}}_{A}\) and \({\bar{c}}_{A}\) denote the curvature and loading of the leave-out proxy. The regression therefore measures the curvature of source \(s\) relative to its leave-out benchmark, which is the object the normalization of Section 2 makes relevant: departures from the consensus scale. A curvature component common to every source is absorbed by the normalization and is not detectable from the sources. The appendix states the result formally. The leading scale factor equals one when the leave-out loadings coincide; after per-source standardization the loadings implied by the observed cross-source correlations differ by a few percent, so the factor is close to one in the application.

This square-root scale correction is important. Without an explicit normalization that sets the leading latent loading of the leave-out proxy equal to one, using \(\operatorname{Cov}(Z_{iA}^{(s)},Z_{iB}^{(s)})\) rather than its square root would leave the curvature estimate on the wrong scale. The covariance itself estimates the product of the two leave-out loadings, while the relative-curvature correction requires the loading of the regressor-side leave-out proxy.

The same auxiliary regression can be augmented with a cubic term as a diagnostic for the third-order remainder in Assumption 4. A cubic coefficient large relative to the squared curvature bound indicates that the second-order approximation may be insufficient over the observed support of the latent variable. In that case, the closed-form interval should be interpreted as a local approximation rather than as a reliable finite-sample summary.

In implementation, we report the identified set~\eqref{eq:idset} using \(\widehat{B}\). We also report a version that replaces \(\widehat{B}\) with a one-sided upper confidence bound. This second interval gives a conservative reference that incorporates sampling uncertainty in the curvature diagnostic.

\subsection{Standardized moments}
The half-width formula also requires the standardized third and fourth moments of the latent regressor. Since the latent regressor is not observed, these moments are estimated from the same cross-source structure used to construct the estimator.

Let

\[{\bar{M}}_{i} = \frac{1}{S}\sum_{s = 1}^{S}M_{i}^{(s)}\]

denote the full cross-source average, after source standardization. Under the measurement model and Assumption 2, this average is a noisy monotone proxy for the latent regressor. The empirical moments used in the half-width are computed from the standardized version of \({\bar{M}}_{i}\). Denote these estimates by \({\widehat{\kappa}}_{3}\) and \({\widehat{\kappa}}_{4}\). Under the normalization \(\sum_{s}\gamma_{s} = 0\) the quadratic terms cancel in the equally weighted average, so \({\bar{M}}_{i}\) is linear in \(E_{i}\) up to the Taylor remainders and the averaged noise, and the plug-in moments estimate the latent moments with an error governed by the noise share of the average rather than by the curvature.

The estimated multiplicative half-width is

\[\widehat{\Delta} = \frac{1}{8}{\widehat{B}}^{2}\left( {\widehat{\kappa}}_{4}-1-{\widehat{\kappa}}_{3}^{2} \right).\]

The estimated identified set is centered at \({\widehat{\beta}}_{sym}\) and uses the multiplicative adjustment implied by the half-width in equation~\eqref{eq:delta}. For small \(\widehat{\Delta}\), the interval is well approximated by

\[\left[ {\widehat{\beta}}_{sym}(1 - \widehat{\Delta}),{\widehat{\beta}}_{sym}(1 + \widehat{\Delta}) \right].\]

The exact endpoint transformation is used in the reported estimates.

\subsection{Partition choice}
The estimator requires a partition of the source set into two disjoint subsets. When there are several admissible partitions, the choice does not change the population content of the identification result, but it can affect finite-sample estimates.

A natural default is to use balanced partitions. With four sources, for example, there are three distinct two-versus-two partitions. If the auxiliary curvature diagnostics indicate similar curvature across sources, no partition is preferred on theoretical grounds. If some sources exhibit systematically larger curvature, a partition that balances estimated curvature across the two halves can reduce finite-sample sensitivity.

The empirical section reports the sensitivity of the estimates to admissible partitions. A further implementation is to average the symmetric estimator across all admissible partitions. This is analogous in spirit to split-sample averaging procedures and reduces dependence on an arbitrary partition choice. The asymptotic theory in Section 5 is stated for a fixed partition, but the same arguments apply to a finite average over admissible partitions after applying the delta method to the enlarged moment vector.

\subsection{Instrument strength and diagnostics}
The split-source construction also yields standard first-stage diagnostics. Under Assumptions 2 and 5, and for a curvature bound small enough that the covariance between the two subset averages is positive, the first-stage coefficient is positive. The first-stage \(F\)-statistic diverges at the parametric rate under the maintained model.

Weak first-stage behavior is therefore informative. It can indicate that the retained sources do not share a sufficiently strong common latent component, that the source partition is poorly chosen, or that the construct-inclusion step has failed to isolate a common factor. For this reason, the empirical implementation reports first-stage slopes and cluster-robust first-stage \(F\)-statistics.

The residual correlation matrix across sources provides a complementary diagnostic for Assumption 2. Large residual correlations across sources after conditioning on the common component would indicate common source-specific errors and would undermine the orthogonality conditions used by both the symmetric estimator and the split-instrument curvature regression.

\subsection{Platform-selection adjustment}
In some applications, source measurements may also reflect selection into the platform or data-generating environment from which the measurement is produced. Appendix B develops an extension in which each source contains an additive platform-selection component. When within-task cross-platform variation is available, this component can be estimated and removed before applying the multi-source estimator.

The main identification and estimation results do not require this adjustment. If the analyst does not observe the data needed to identify platform selection separately, the adjustment can be omitted. In that case, any platform-specific component is absorbed into the source-specific measurement error. The validity of the baseline estimator then depends on whether the resulting error satisfies the orthogonality restrictions in Assumption 2.

\section{Asymptotic Theory and Inference}
This section gives the large-sample theory for the estimated identified set and the associated confidence interval. Section 5.1 defines the moment vector and establishes joint asymptotic normality of the estimated endpoints. Section 5.2 constructs a uniformly valid confidence interval using the Imbens and Manski procedure with the Stoye critical value. Section 5.3 discusses diagnostics for the maintained restrictions.

\subsection{Endpoint asymptotics}
The estimated identified set depends on three objects: the symmetric estimator, the estimated curvature bound, and the estimated standardized moments of the latent regressor. These objects can be written as smooth transformations of a common moment vector. Let

\[\widehat{m} = \left( {\widehat{\sigma}}_{Y1},{\widehat{\sigma}}_{Y2},{\widehat{\sigma}}_{12},\{{\widehat{b}}_{1s},{\widehat{b}}_{2s},{\widehat{\sigma}}_{AB,s}\}_{s = 1}^{S},{\widehat{\kappa}}_{3},{\widehat{\kappa}}_{4} \right),\]

where \({\widehat{\sigma}}_{Y1}\), \({\widehat{\sigma}}_{Y2}\), and \({\widehat{\sigma}}_{12}\) are the covariance estimates entering the symmetric estimator, \({\widehat{b}}_{1s}\) and \({\widehat{b}}_{2s}\) are the coefficients from the split-instrument auxiliary regression for source \(s\), \({\widehat{\sigma}}_{AB,s}\) is the covariance between the two leave-out averages used in that auxiliary regression, and \({\widehat{\kappa}}_{3}\) and \({\widehat{\kappa}}_{4}\) are the estimated standardized third and fourth moments.

Under Assumption 5, the components of \(\widehat{m}\) are sample averages, or smooth functions of sample averages, with finite second moments implied by the eighth-moment condition. Hence,

\begin{equation}
\sqrt{n}(\widehat m - m) \xrightarrow{d} \mathcal{N}(0, \Omega),
\end{equation}

where \(\Omega\) is consistently estimable. Under independent sampling, \(\Omega\) is estimated by the usual plug-in covariance matrix of the empirical moment functions. Under clustered sampling, it is estimated by the corresponding cluster-robust covariance matrix.

Let

\[\widehat{\Delta} = \frac{1}{8}{\widehat{B}}^{2}\left( {\widehat{\kappa}}_{4}-1-{\widehat{\kappa}}_{3}^{2} \right)\]

denote the estimated multiplicative half-width, and let \({\widehat{\beta}}_{sym}\) denote the symmetric estimator defined in equation~\eqref{eq:symest}. The estimated endpoints are

\[{\widehat{\theta}}_{l} = \min\{{\widehat{\beta}}_{sym}(1 - \widehat{\Delta}),{\widehat{\beta}}_{sym}(1 + \widehat{\Delta})\},\]

and

\[{\widehat{\theta}}_{u} = \max\{{\widehat{\beta}}_{sym}(1 - \widehat{\Delta}),{\widehat{\beta}}_{sym}(1 + \widehat{\Delta})\}.\]

The orientation by minimum and maximum is necessary because the structural coefficient may be negative.

\textbf{Proposition 2. Endpoint asymptotics.}\\
Suppose Assumptions 1 through 6 hold. Suppose also that \(\sigma_{12} > 0\), that the denominator term in Proposition 1 is bounded away from zero, and that the endpoint variances are positive. Then

\[\begin{aligned}
\sqrt{n}\left[ \left( \begin{array}{r}
{\widehat{\theta}}_{l} \\
{\widehat{\theta}}_{u}
\end{array} \right)-\left( \begin{array}{r}
\theta_{l} \\
\theta_{u}
\end{array} \right) \right] \xrightarrow{d} N(0,\Sigma_{\theta}),
\end{aligned}\]

where

\[\Sigma_{\theta} = J(m)\Omega J(m)'.\]

Here \(J(m)\) is the Jacobian of the endpoint map \(m \mapsto (\theta_{l},\theta_{u})\). The population endpoints \(\left( \theta_{l},\theta_{u} \right)\) correspond to the second-order identified set in Theorem 1, up to the \(O(B^{3})\) approximation error.

Assumption 6 ensures that the maximum defining \(B\) is locally differentiable. If the maximum is attained by more than one source, the same conclusion can be obtained using directional differentiability arguments, although the limiting distribution is no longer represented by an ordinary Jacobian.

For clustered data, the covariance matrix \(\Omega\) is estimated by replacing the individual outer products of moment-function deviations with cluster-level outer products. The empirical application clusters at the occupation level. When the number of clusters is small, a wild-cluster bootstrap applied to the influence functions of the full moment vector provides a useful robustness check. The bootstrap should be applied to the joint endpoint vector, rather than to the point estimator alone, so that sampling variation in the estimated half-width is preserved.

\subsection{Uniform confidence intervals}
The parameter \(\beta\) belongs to the population interval \(\left[ \theta_{l},\theta_{u} \right]\), up to the approximation error characterized in Theorem 1. Inference therefore concerns a scalar parameter lying in an estimated interval. We use the confidence interval construction of Imbens and Manski (2004), with the critical value adjustment of Stoye (2009), which gives uniform coverage as the identified set shrinks to a point.

Let \({\widehat{\sigma}}_{l}^{2}\) and \({\widehat{\sigma}}_{u}^{2}\) denote the diagonal elements of \({\widehat{\Sigma}}_{\theta}\). These are the asymptotic variances of \(\sqrt{n}({\widehat{\theta}}_{l} - \theta_{l})\) and \(\sqrt{n}({\widehat{\theta}}_{u} - \theta_{u})\). The corresponding standard errors are \({\widehat{\sigma}}_{l}/\sqrt{n}\) and \({\widehat{\sigma}}_{u}/\sqrt{n}\).

Define \(c_{n}\) as the solution to

\begin{equation}
\Phi\!\left(c_n + \frac{\sqrt{n}\,(\hat\theta_u - \hat\theta_l)}{\max(\hat\sigma_l, \hat\sigma_u)}\right) - \Phi(-c_n) = 1 - \alpha, \label{eq:im}
\end{equation}

The \(1 - \alpha\) confidence interval is

\begin{equation}
C_n(1-\alpha) = \left[\,\hat\theta_l - c_n\frac{\hat\sigma_l}{\sqrt{n}},\ \hat\theta_u + c_n\frac{\hat\sigma_u}{\sqrt{n}}\,\right]. \label{eq:ci}
\end{equation}

\textbf{Theorem 2. Uniformly valid inference.}\\
Let \(\mathcal{P}(B_{\max})\) denote the class of data-generating processes satisfying Assumptions 1 through 6 with relative curvature bounded by \(B_{\max}\), uniformly bounded eighth moments, \(\sigma_{12}\) and \(\lvert \beta \rvert\) bounded away from zero, endpoint variances bounded away from zero and infinity, and the curvature gap in Assumption 6 bounded away from zero. Consider sequences \(B_{\max} = B_{n}\) with \(\sqrt{n}\,B_{n}^{3} \rightarrow 0\). Then

\begin{equation}
\liminf_{n\to\infty}\,\inf_{P \in \mathcal{P}(B_{n})} P\{\beta \in C_n(1-\alpha)\} \geq 1 - \alpha.
\end{equation}

The rate condition \(\sqrt{n}\,B_{n}^{3} \rightarrow 0\) reflects the second-order nature of the identified set in Theorem 1. The set is constructed from a second-order expansion, so the structural parameter can lie at a distance of order \(B^{3}\) from the population endpoints, while the sampling margins of the confidence interval contract at rate \(n^{-1/2}\). The condition keeps the approximation error negligible relative to those margins. At a fixed curvature bound the statement holds for the second-order identified set itself, or for the structural parameter after the interval is widened by the third-order remainder bound.

The Stoye critical value is important because the width of the identified set may be small. When the population width is large relative to sampling error, the procedure behaves like inference on a partially identified parameter. When the width is zero, it reduces to the point-identified case. The construction therefore avoids the discontinuity that arises from using a fixed critical value that is appropriate only in one of the two regimes.

The proof verifies the conditions required for the Imbens-Manski-Stoye interval~\eqref{eq:ci}: joint asymptotic normality of the estimated endpoints, consistent estimation of the endpoint variances, and consistent estimation of the interval width. Proposition 2 supplies the first condition. The plug-in or cluster-robust covariance estimator supplies the second. The continuous mapping theorem and Assumption 6 supply the third.

\subsection{Diagnostics for maintained restrictions}
The validity of the confidence interval depends on the maintained measurement restrictions. Two diagnostics are especially important.

First, the split-instrument curvature diagnostic should support the second-order approximation. The auxiliary regression in Section 4 estimates the quadratic component of each measurement function relative to the consensus scale. A cubic extension of the auxiliary regression provides a diagnostic for the third-order remainder. A large and statistically significant cubic term would indicate that the closed-form interval may not summarize the relevant curvature bias accurately over the observed support.

Second, the source-level errors should not exhibit strong residual dependence after accounting for the common latent component. This condition is assessed using the residual correlation matrix across sources. Large residual correlations would violate the cross-source orthogonality restrictions in Assumption 2 and could affect both the symmetric estimator and the split-instrument curvature diagnostic.

If these diagnostics fail, the interpretation of the interval should change. The interval can still be reported as a sensitivity calculation under the stated curvature and orthogonality restrictions, but it should not be presented as the baseline identified set. Possible responses include restricting attention to source families with weaker residual dependence, modeling common residual factors explicitly, or treating the curvature bound as an externally specified sensitivity parameter.

The scale of the latent regressor is fixed by the normalization in Assumption 1. In applications with occupation-level measurements, source standardization should be performed on the analysis sample used in the downstream regression. Standardizing on a broader universe changes the numerical scale of the exposure variable and therefore changes the interpretation of \(\beta\). The empirical application uses the occupation set in the analysis sample for this reason.

\section{Monte Carlo Evidence}
This section reports Monte Carlo evidence on the finite-sample behavior of the estimator and confidence interval. The simulations have three goals. First, they assess whether the bias of the symmetric estimator is bounded by the second-order identified set derived in Theorem 1. Second, they examine the accuracy of the delta-method standard errors in Proposition 2. Third, they evaluate the coverage of the Imbens-Manski confidence interval with the Stoye critical value in Theorem 2.

\subsection{Design}
The baseline data-generating process is

\[Y_{i} = \beta E_{i} + \varepsilon_{i},\beta = 1,\]

where \(\varepsilon_{i}\) is Gaussian and independent of the latent exposure \(E_{i}\). The latent exposure is drawn from three standardized distributions. The first is Gaussian, corresponding to the benchmark in Corollary 2. The second is Student-\(t\) with five degrees of freedom, rescaled to have mean zero and variance one, which introduces excess kurtosis with little skewness. The third is a centered and standardized chi-squared distribution with three degrees of freedom, which introduces both skewness and excess kurtosis.

For each replication, four noisy measurements of \(E_{i}\) are generated according to

\[M_{i}^{(s)} = c_{s}E_{i} + \frac{1}{2}\gamma_{s}E_{i}^{2} + U_{i}^{(s)},s = 1,\ldots,4,\]

with independent Gaussian source errors. The relative-curvature bound takes values between 0.05 and 0.40, under both alternating-sign and same-sign curvature patterns, and the sample size takes four values between 1,000 and 125,000. Each design cell is replicated 500 times.

\begin{figure}[h!]
\centering
\includegraphics[width=0.95\linewidth,keepaspectratio]{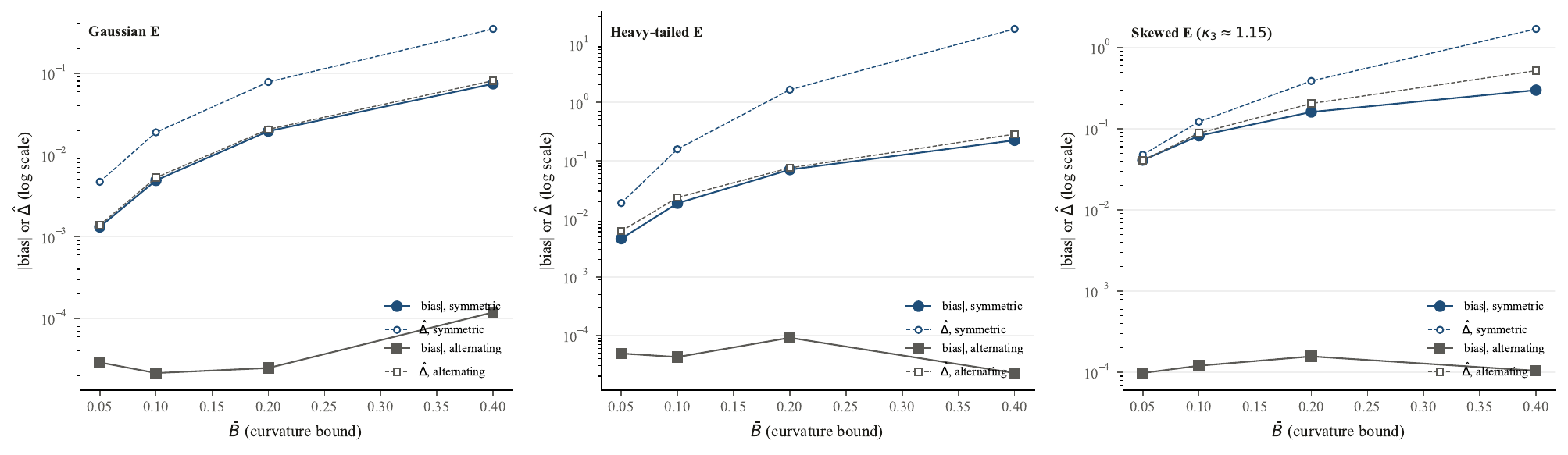}
\caption{Monte Carlo verification of Theorem 1 across designs.}\label{fig:2}
\begin{minipage}{0.95\linewidth}
\footnotesize\textbf{Notes.} Empirical bias and Theorem 1 half-width $\hat\Delta$ from 48,000 simulated fits across 96 design cells consisting of three latent distributions (Gaussian, heavy-tailed, skewed), four curvature bounds ($\bar B \in \{0.05, 0.10, 0.20, 0.40\}$), two curvature patterns (alternating and symmetric), and four sample sizes ($n \in \{1{,}000, 5{,}000, 25{,}000, 125{,}000\}$). The bias never exceeds $\hat\Delta$ in any cell, and the combined-inference coverage of Theorem 2 exceeds 95\% in every cell.
\end{minipage}
\end{figure}

For each replication, we compute the symmetric estimator, the half-width evaluated at the design curvature bound, the endpoint standard errors, and the Imbens-Manski-Stoye confidence interval. The grid therefore isolates the behavior of the estimator and the interval at a known bound. The fully feasible procedure, in which the curvature bound and the latent moments are estimated within each replication, is examined in the calibrated design below. Results across the design cells are summarized in Figure~\ref{fig:2}.

\subsection{Baseline results}
The baseline simulations support the main theoretical predictions. At the largest sample size, the absolute bias of the symmetric estimator is below the Theorem 1 half-width in every design cell~(Figure~\ref{fig:2}). The bound is closest to binding when the latent distribution is skewed and the curvature pattern is same-signed, which is the configuration in which the second-order curvature term is most pronounced.

The delta-method standard errors from Proposition 2 track the simulation standard deviations closely in the larger samples. Discrepancies concentrate in the skewed, high-curvature designs, where the score distribution is heavy-tailed and the normal approximation less accurate.

The Imbens-Manski-Stoye interval attains at least nominal coverage in the simulation grid. Coverage is conservative in many cells, especially when the identified set is wide relative to sampling uncertainty. The procedure also behaves appropriately near the point-identified boundary. When \(B = 0\), the interval reduces smoothly to the point-identified case, while away from the boundary it behaves as an interval for a partially identified parameter.

\subsection{Loading heterogeneity}
The first additional design studies loading invariance. Source loadings are drawn from a heterogeneous distribution, the latent exposure is skewed, and source curvatures vary across measurements, a configuration that isolates the difference between the symmetric estimator and the asymmetric two-stage least-squares estimator.

The symmetric estimator remains centered near the structural coefficient. The asymmetric estimator does not: its probability limit depends on the loading of the endogenous measurement, and it displays sizeable bias in this design. The contrast illustrates how the symmetric covariance product eliminates unknown source-specific loadings, and it argues for treating the asymmetric estimator as a robustness specification unless stronger loading restrictions are imposed.

\subsection{Correlated source errors}
The second additional design varies the correlation across source-level measurement errors. This design directly probes Assumption 2. As expected, the symmetric estimator is stable when cross-source error correlations are small, but its bias increases as the correlation rises. The reason is that common source-level noise affects the covariance between subset averages, which enters the denominator of the symmetric estimator.

This exercise clarifies the interpretation of the residual-correlation diagnostic used in the empirical application. Small residual correlations across sources support the orthogonality restrictions used by the estimator. Large residual correlations would indicate that the baseline identified set should be interpreted as conditional on a violated restriction, or that the analyst should model common residual factors explicitly.

\subsection{Split-instrument curvature estimation}
The third additional design evaluates the split-instrument curvature diagnostic. When the true relative curvature is zero, the split-instrument auxiliary regression produces estimates centered close to zero, which confirms that the orthogonality of the two leave-out averages generates no spurious curvature. When the same curvature is imposed on every source, the split-instrument estimates are again centered close to zero: a component common to all sources is absorbed by the latent normalization of Section 2 and is not detectable from the sources. The diagnostic has power against curvature heterogeneity, which is the object that enters the identified set under the normalization.

The direct OLS version of the auxiliary regression does neither. It produces nonzero curvature estimates even at zero true curvature, because the leave-out proxy is measured with error, and under common curvature its output mixes that errors-in-variables contamination with the component the sources cannot reveal. These patterns support using the split-instrument version, interpreted as an estimate of curvature relative to the consensus scale, for the bound entering the identified set.

\subsection{Calibrated design}
The calibrated design matches the empirical application in both respects. The latent regressor is drawn from a two-point location mixture with Gaussian noise whose third and fourth standardized moments are 0.180 and 1.690, against 0.183 and 1.649 in the four-language-model panel, and the source curvatures are the departures implied by the panel's split-instrument contrasts under the normalization, so the maximal leave-out contrast in the design sits close to the empirical estimate. Nothing is treated as known within a replication: the curvature bound is estimated by the per-source split-instrument regression and the moments by the standardized cross-source average. The mean estimated bound is 0.391. The mean feasible half-width is 1.46 percent, and the estimation error of the symmetric estimator falls inside the feasible band in 99.8 percent of replications. The cubic extension of the auxiliary regression produces a mean coefficient of 0.065, within the order permitted by the third-derivative restriction in Assumption 4 at the estimated bound.

The feasible half-width in this design is close to the empirical value of 1.47 percent for the four-language-model panel, which is the intended calibration. The second-order identification bias implied by the subset curvatures is an order of magnitude smaller than the feasible band, so most of the estimation error inside the band is sampling variation rather than curvature bias. The slack arises because the subset averages damp offsetting departures across sources.

\subsection{Scope implied by the simulations}
The simulations also indicate when the closed-form interval becomes less informative. With high curvature and heavy-tailed latent exposure, the estimated half-width can become large relative to the point estimate. In such designs, the interval remains valid as a second-order bound, but it may provide limited empirical content. This is not a failure of the inference procedure. It reflects the fact that substantial nonlinear measurement distortion and heavy-tailed latent variation leave little information about the structural coefficient without stronger restrictions.

The simulations are intentionally stylized. They impose the cross-source orthogonality condition by construction, omit platform-selection adjustments, and use independent sampling rather than clustered data. These simplifications are appropriate for isolating the theoretical mechanisms, but they also identify the diagnostics that matter in applications: residual dependence across sources, the magnitude of the cubic term in the auxiliary regression, and the robustness of standard errors to clustering.

\section{Empirical Application}
This section applies the proposed method to occupational measures of exposure to artificial intelligence. The application has two purposes. First, it documents the extent to which single-source exposure measures produce different downstream labor-market coefficients when applied to the same data. Second, it shows how the multi-source framework reconciles this disagreement by separating distinct latent constructs, estimating the curvature bound, and constructing a partial-identification interval around a loading-invariant consensus estimate.

The application should be interpreted as a measurement-reconciliation exercise. Standard parallel-trends diagnostics reject the maintained design for the full post-2022 difference-in-differences specification, largely because of the COVID-era composition reversal in 2020 and 2021. We therefore report the full-sample estimates, COVID-excluded estimates, and event-study diagnostics, but we do not attach a causal displacement interpretation to the main post-2022 coefficient.

\subsection{Data}
The empirical analysis combines three data sources. The first is a multi-source occupation-level panel of AI-exposure measures aggregated to six-digit Standard Occupational Classification codes. It contains six exposure measures. Four are large-language-model ratings of O*NET task descriptions under the rubric of Eloundou, Manning, Mishkin, and Rock (2024), generated by ChatGPT-5, Claude 4.5, GPT-4 in the original Eloundou specification, and Gemini 2.5. The remaining two are the ability-weighted exposure index of Felten, Raj, and Seamans (2021) and the patent-text matching score of Webb (2020).

The second data source is the 2015 to 2024 American Community Survey individual-level extract from IPUMS USA. The sample is restricted to civilian wage-and-salary workers aged 16 to 64. The analysis sample contains 8,883,132 person-year observations across 379 six-digit SOC codes, corresponding to the intersection of the language-model-rated occupations and the occupations observed in the ACS extract.

The third data source is used for validation. We use the 2024 Real-Time Population Survey AI module (Bick, Blandin, Deming, and Schumacher, 2026), which provides worker-reported at-work AI-use rates at the six-digit SOC level for 333 occupations overlapping the analysis sample. We also construct an LLM-free O*NET anchor from the first principal component of 232 O*NET attributes covering skills, abilities, knowledge, work activities, work styles, and work context.

\subsection{Construct inclusion and the Webb measure}
The six exposure measures need not correspond to a single latent construct. This issue is central in the present application because the patent-text measure of Webb (2020) differs both conceptually and empirically from the language-model-based measures. The language-model measures and the Felten et al. index are designed to capture the current applicability of AI to occupational tasks or abilities. The Webb measure instead captures overlap between occupational task descriptions and AI-related patent text, which is closer to a measure of technological substitutability by inventive activity.

We therefore impose an ex ante construct-inclusion rule. A source enters the multi-source construction only if its per-source Kaiser-Meyer-Olkin sampling adequacy exceeds 0.6 and Horn's parallel-analysis criterion retains it on the single shared factor. This rule is applied before downstream labor-market coefficients are used, so the inclusion decision is not outcome-dependent.

\begin{figure}[h!]
\centering
\includegraphics[width=0.95\linewidth,keepaspectratio]{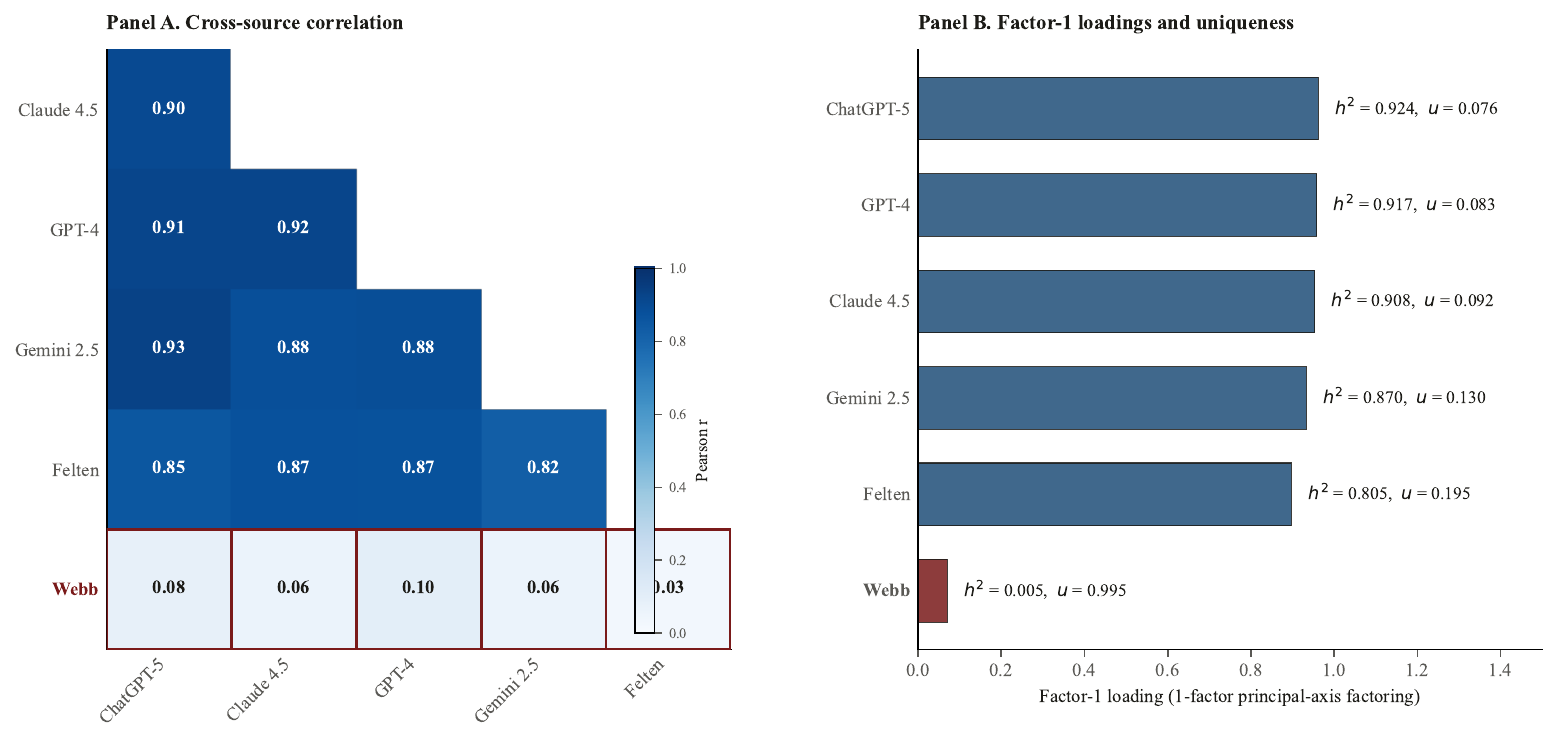}
\caption{Webb (2020) is a separate latent construct from the LLM-and-Felten group.}\label{fig:1}
\begin{minipage}{0.95\linewidth}
\footnotesize\textbf{Notes.} Panel A reports the lower-triangle Pearson correlation matrix among the six AI-exposure measures on the 682-occupation SOC6 panel. The four LLM raters and Felten (2021) are mutually correlated above 0.82, while Webb's row (red border) shows correlations below 0.11 with every other source. Panel B reports the factor-1 loadings and uniqueness from a 1-factor principal-axis factor analysis on the six-source panel. The LLM-and-Felten group loads 0.90 to 0.96 on the shared factor with uniqueness between 0.076 and 0.195, while Webb loads 0.07 with uniqueness 0.995. The two panels jointly identify Webb as a separate latent construct, so aggregation across the LLM-Felten union and Webb is unwarranted.
\end{minipage}
\end{figure}

The six-source correlation matrix (Table~\ref{tab:t1}) passes a broad common-factor diagnostic but reveals that Webb is not part of the same construct. Bartlett's test of sphericity rejects the identity correlation matrix, and the overall Kaiser-Meyer-Olkin statistic is 0.904. The four language-model measures and the Felten index have per-source sampling adequacy between 0.875 and 0.943. The Webb measure has per-source sampling adequacy 0.453, below the inclusion threshold. In the one-factor principal-axis solution (Figure~\ref{fig:1}), the four language-model measures and Felten load between 0.90 and 0.96, with uniqueness below 0.20. Webb loads at 0.07, with uniqueness 0.995.

We therefore use the four language-model measures and the Felten index as the five-source panel for the baseline multi-source estimator. The Webb measure is reported separately as an alternative construct. This distinction is important for interpretation: disagreement between Webb and the other measures is not treated as measurement error around a common latent object, but as evidence that the patent-text measure captures a different exposure concept.

\subsection{Single-source estimates}
The downstream specification relates worker outcomes to the interaction between occupational AI exposure and a post-2022 indicator, controlling for occupation, state, and year fixed effects and standard worker-level covariates. The outcomes are employment, labor-force participation, and log wages.

\begin{figure}[h!]
\centering
\includegraphics[width=0.95\linewidth,keepaspectratio]{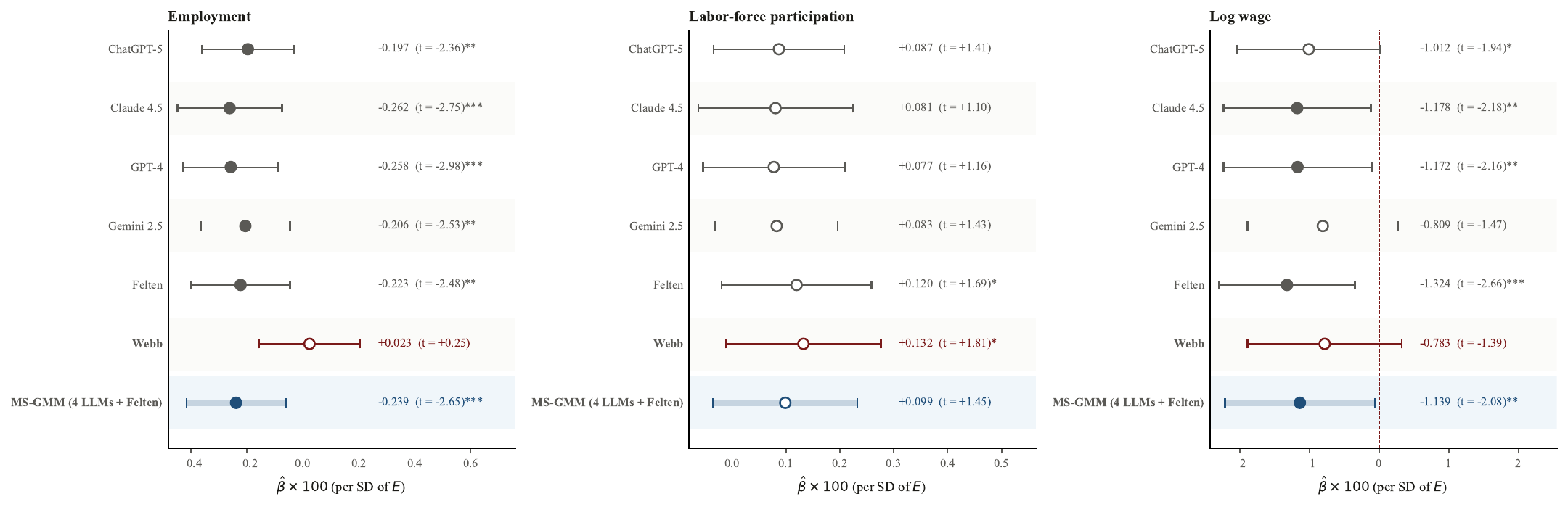}
\caption{Single-source vs MS-GMM downstream estimates, three labor-market outcomes.}\label{fig:3}
\begin{minipage}{0.95\linewidth}
\footnotesize\textbf{Notes.} Forest plot of post-2022 difference-in-differences coefficients on the indicated outcome for each of seven estimators, estimated on the 8.88-million-worker American Community Survey panel. Single-source estimators treat each AI-exposure measure (ChatGPT-5, Claude 4.5, GPT-4, Gemini 2.5, Felten, Webb) as a noiseless proxy for the latent exposure. The MS-GMM 4 LLMs + Felten estimator is the cross-source half-average two-stage least squares estimator of Section 3. Filled markers denote significant coefficients at the 5\% level; hollow markers denote nonsignificant coefficients. The factor-of-eleven spread across single-source employment elasticity estimates (from $+0.023$ on Webb to $-0.262$ on Claude 4.5) is reconciled by the MS-GMM consensus estimate at $-0.239$.
\end{minipage}
\end{figure}

Across the six single-source exposure measures, the post-2022 employment coefficient differs substantially, ranging from \(- 0.262\) for Claude 4.5 to \(+ 0.023\) for Webb~(Table~\ref{tab:t3}; Figure~\ref{fig:3}). The four language-model measures and Felten produce negative and statistically significant employment coefficients in the range \([ - 0.262, - 0.197]\). The Webb estimate is positive and statistically indistinguishable from zero.

This dispersion is the empirical motivation for the method. If each source were an interchangeable linear measure of the same latent variable, the downstream estimates would differ mainly by scale. Instead, the estimates differ in magnitude and, for Webb, in sign. The multi-source procedure below first separates Webb as a distinct construct and then estimates a loading-invariant consensus coefficient using the five retained sources.

\subsection{Baseline multi-source estimate}
Applying the symmetric cross-source estimator to the five-source language-model-and-Felten panel yields a post-2022 employment coefficient of \(- 0.239\), with a cluster-robust standard error of 0.090. The estimate lies within the range of the five retained sources and is separated from the Webb single-source estimate~(Figure~\ref{fig:a3}).

The partial-identification half-width~\eqref{eq:delta} implied by Theorem 1 is 1.23 percent of the point estimate at the estimated curvature bound, or 1.88 percent when the bound is replaced by its one-sided 95 percent upper confidence limit. The combined inference region, which applies the Imbens-Manski construction with the Stoye critical value to the estimated band endpoints, is \([ - 0.419, - 0.064]\)~(Table~\ref{tab:t3}).

Thus, in this application, the residual nonlinear-measurement uncertainty around the five-source consensus is small relative to sampling uncertainty. The main empirical issue is not the remaining curvature bias within the retained construct but the choice between empirically distinct exposure constructs, above all the distinction between Webb's patent-text measure and the language-model-and-Felten group.

\subsection{Curvature diagnostic and worked implementation}
The bounded-curvature assumption is disciplined by the split-instrument auxiliary regression~\eqref{eq:aux} described in Section 4. For each source, we regress the standardized source on a leave-out source average and its square, instrumenting with a disjoint leave-out average and its square. The estimated relative curvature of source \(s\) is obtained from the IV coefficient on the squared proxy and the corresponding first-stage scale adjustment.

The diagnostic rejects linearity for every language-model measure at the 1 percent level. The estimated relative-curvature bound from equation~\eqref{eq:curv} is \(B = 0.4380\) on the four-language-model panel and \(B = 0.4147\) on the five-source panel that adds Felten~(Table~\ref{tab:t2}). In both panels, the maximum split-instrument curvature estimate is attained by ChatGPT-5. Under the normalization of Section 2 these estimates measure curvature relative to the consensus scale, and \(B\) bounds the heterogeneity of the measurement functions across sources rather than their common shape.

The OLS version of the same RESET-style diagnostic gives smaller bounds, 0.344 on the four-language-model panel and 0.273 on the five-source panel~(Figure~\ref{fig:a1}). This difference in the bound is consistent with the errors-in-variables attenuation described in Section 4: because the leave-out average is itself noisy, the OLS auxiliary regression understates the curvature bound, defined as the maximum absolute relative curvature across raters. The correction acts on this maximum rather than uniformly rater by rater; the OLS and split-instrument diagnostics use different leave-out proxies, so individual raters can move in either direction, but the bound that enters the identified set is larger under the split-instrument construction. 

The standardized moments of the cross-source average are \(\kappa_{3} = 0.183\) and \(\kappa_{4} = 1.649\) on the four-language-model panel, and \(\kappa_{3} = 0.133\) and \(\kappa_{4} = 1.591\) on the five-source panel. Substituting these estimates into the symmetric half-width formula gives half-widths of 1.47 percent for the four-language-model panel and 1.23 percent for the five-source panel. Using the one-sided 95 percent upper confidence bound for \(B\) raises these values to 2.47 percent and 1.88 percent, respectively~(Table~\ref{tab:t2}).

The four-language-model panel provides a useful worked implementation. Under the balanced partition placing ChatGPT-5 and Claude 4.5 in one subset and GPT-4 and Gemini 2.5 in the other, the first-stage slope is 0.996 and the first-stage \(F\)-statistic is 2,981. The downstream symmetric estimate is \(- 0.264\), with a cluster-robust standard error of 0.088~(Table~\ref{tab:t4}). This four-language-model estimate lies just outside the range of the four single-source coefficients, \([ - 0.262, - 0.197]\). This is expected rather than anomalous: each single-source regression is attenuated toward zero by measurement error, so the loading-invariant estimator, which removes that attenuation, can be larger in magnitude than any individual source. Adding the Felten index in the five-source panel pulls the consensus to \(- 0.239\), inside the retained-source range. The symmetric and asymmetric estimators are numerically close in this case because the first-stage slope is near one, but only the symmetric estimator is invariant to unknown source loadings in the general model.

\subsection{Partition sensitivity and first-stage strength}
The source split used to form the half-averages can affect finite-sample estimates, even though the asymptotic identification argument is invariant to the choice of a fixed admissible partition. We therefore report sensitivity to all admissible balanced partitions of the four-language-model source set.

Across the three admissible two-versus-two partitions, the cross-source estimate moves by no more than \(\pm 0.0001\) in the employment specification, negligible relative to the cluster-robust standard error of 0.090~(Table~\ref{tab:t3}). The Theorem 1 half-width changes by less than 0.001 across partitions.

The first-stage diagnostics also support the retained source set: in the main four-language-model implementation the first-stage slope is 0.996 with an \(F\)-statistic of 2,981, and across the subgroup-outcome cells reported later in the section and in the appendix the first-stage \(F\)-statistics range from 600 to 3,700. The retained sources share a strong common component.

\subsection{Residual dependence and an LLM-free O*NET anchor}
Assumption 2 requires that source-specific measurement errors not exhibit strong residual dependence after conditioning on the latent exposure. This restriction is particularly relevant for large language models because different models may share training data, task descriptions, or annotation conventions. We assess this issue in two ways.

First, we examine residual correlations across sources after fitting the common component. In the four-language-model panel the off-diagonal residual correlations are below 0.1 in absolute value for every pair, against raw correlations above 0.88~(Table~\ref{tab:t1}), which supports the cross-source orthogonality approximation in the baseline implementation. 

Second, we construct an LLM-free validation anchor from O*NET features: the first principal component of 232 O*NET attributes observed on the O*NET occupation frame, unsupervised and using no AI-exposure score. The first component explains 32.6 percent of feature variance, with eigenvalue 75.5. Its positive loadings fall on cognitive and communication-intensive features, including written comprehension, reading comprehension, programming, electronic-mail use, and working with computers; its negative loadings fall on physical features, including manual dexterity, static strength, multilimb coordination, extent flexibility, and repairing.

The O*NET anchor correlates strongly with the retained AI-exposure construct. Its Spearman correlation with the cross-source-corrected exposure score is 0.868, with the four-language-model-only construction is 0.840, and with the Felten index is 0.914. Its correlation with the Webb patent-text measure is \(- 0.071\)and statistically weak. This pattern reinforces the factor-analytic result~(Figure~\ref{fig:1}): the language-model-and-Felten group captures an O*NET-based cognitive-task construct, while Webb captures a different object.

The anchor is also stable. Dropping any one of the six O*NET domains and re-extracting the first principal component yields a Spearman correlation of at least 0.98 with the full-feature component. Bootstrap resampling of occupations gives fifth-percentile rank correlation above 0.998 with the original component.

\subsection{External validation with worker-reported AI use}
We next compare the corrected exposure score with worker-reported AI use from the 2024 Real-Time Population Survey AI module. On the 333 overlapping six-digit SOC codes, the corrected score has Spearman correlation 0.28 with worker-reported AI-use rates. The correlation is statistically distinguishable from zero at conventional levels. The Pearson correlation is 0.22.

The magnitudes are moderate, as expected. Task-level applicability and realized worker adoption are distinct constructs. A technology may be applicable to an occupation without being widely adopted, and workers may use AI tools in ways not fully captured by task-based exposure scores. The positive relationship nevertheless provides external validation that the corrected score captures a component of observed occupational AI use.

\subsection{Parallel trends and COVID-excluded estimates}
The post-2022 difference-in-differences design does not pass standard parallel-trends diagnostics in the full 2015 to 2024 sample. Joint tests on pre-2022 year-by-exposure interactions reject for employment, labor-force participation, and log wages. The rejection is concentrated in 2020 and 2021, when high-exposure occupations gained employment as professional work shifted toward remote arrangements during the COVID period.

The event-study coefficients show this reversal clearly~(Figures~\ref{fig:a4} and~\ref{fig:a6}). The employment year-by-exposure estimates for 2020 and 2021 are 0.94 and 0.56, respectively, compared with a pre-2020 average of \(- 0.35\). Excluding the COVID years changes the aggregate post-2022 employment coefficient from a negative full-sample estimate to a statistically insignificant estimate of 0.042, with standard error 0.114. The corresponding COVID-excluded labor-force participation coefficient is 0.300, with standard error 0.083, and the log-wage coefficient is \(- 0.857\), with standard error 0.577.

These results support a cautious interpretation. The full-sample employment coefficient should not be read as a causal estimate of AI-induced displacement. The empirical application instead demonstrates how multiple exposure measures can be reconciled, how distinct constructs can be separated, and how residual nonlinear-measurement uncertainty can be quantified.

\subsection{Subgroup analysis}
We also report subgroup estimates~(Figure~\ref{fig:a5}) to illustrate how the corrected exposure score behaves across worker groups. The most consistent pattern appears for workers with disabilities, identified by the four-domain ACS definition, a group accounting for 5.92 percent of the analysis sample.

For workers with disabilities, the post-2022 employment coefficient is 0.888 in the full sample and 1.101 in the COVID-excluded sample. Both estimates are significant at the 1 percent level. The corresponding labor-force participation coefficients are 1.297 and 1.626. The wage coefficients are \(- 1.903\) and \(- 2.032\).

This pattern is consistent with the interpretation that AI tools may function as workplace accommodations for some workers with mobility, cognitive, self-care, or independent-living limitations. This interpretation should remain cautious because the design is not causal. The disability subgroup is the only positive-employment-coefficient subgroup that survives the Benjamini-Hochberg false-discovery-rate correction at the five percent level across the 153 subgroup-outcome combinations reported in the appendix.

\subsection{Alternative multi-rater correction frameworks}
The discrete-limit interpretation in Sections 2 and 3 links the continuous measurement framework to latent-class and multi-rater correction methods. We therefore compare the proposed estimator with Dawid-Skene maximum likelihood, the obviously-related-IV estimator, and an Imbens-Manski partial-identification region applied to the four-language-model panel.

The estimated rater attenuation factors for the four language models range from 0.775 for Gemini 2.5 to 0.850 for ChatGPT-5~(Figure~\ref{fig:a7}), a narrow band consistent with the tight cluster of single-rater estimates among the language-model measures.

The four correction frameworks produce similar empirical conclusions~(Table~\ref{tab:t4}). Dawid-Skene maximum likelihood, ORIV, the Imbens-Manski region, and the MS-GMM estimator agree in sign and broad magnitude for employment and wages. Their sampling confidence intervals overlap across the three outcomes. The Imbens-Manski region contains the MS-GMM point estimate for every outcome, including labor-force participation, where the MS-GMM estimate is 0.077 and the Imbens-Manski region is \([ - 0.047,0.092]\).

The Dawid-Skene and ORIV point estimates need not fall inside the narrow MS-GMM curvature band, because that band is not a sampling confidence interval and the estimators rely on different identifying restrictions. The relevant empirical finding is that distinct correction frameworks, one based on continuous bounded-curvature measurement and the others based on discrete-rater correction, deliver compatible downstream conclusions in this application.

\section{Released Scores and an O*NET-Based Own-Measure}
This section describes two companion outputs. The first is an occupation-level corrected exposure score constructed from the five-source language-model-and-Felten panel. The score is released with the partial-identification band implied by Theorem 1. The second is an O*NET-based own-measure that approximates the corrected score without using large-language-model output. The two objects serve different purposes. The corrected score is the direct empirical implementation of the multi-source estimator. The own-measure is a sparse, transparent proxy for researchers who prefer a measure constructed only from conventional O*NET attributes.

\subsection{Corrected exposure score}
The corrected exposure score is constructed for the 682 six-digit Standard Occupational Classification codes with full source coverage~(Figure~\ref{fig:a2}). For each occupation, the released table reports the five-source corrected exposure score based on the language-model-and-Felten construct defined in Section 7. The table also reports the lower and upper endpoints of the Theorem 1 partial-identification interval, evaluated at the empirical symmetric half-width. 

The score table includes several additional fields. It reports an alternative four-language-model-only score, the Webb (2020) patent-text exposure measure as a separate construct, and the Bick et al. worker-reported AI-use rate where available. It also includes metadata on source coverage and release version. These fields allow downstream users to distinguish the current-generation applicability construct from the patent-text construct and to track the version of the score used in empirical work. 

The released partial-identification band is multiplicative. A downstream researcher can therefore propagate the band into a regression coefficient by applying the same proportional adjustment to the coefficient obtained with the corrected score. Sampling uncertainty from the downstream regression should then be combined with the partial-identification width using the inference procedure described in Section 5.

The Webb measure is not included in the baseline corrected score. As shown in Section 7.2, the patent-text measure fails the ex ante construct-inclusion rule and behaves as a distinct latent construct. It is therefore reported separately. Applications concerned with current-generation AI applicability should use the corrected language-model-and-Felten score. Applications concerned with patent-based technological substitutability may instead use the Webb measure, but it should not be interpreted as another noisy source for the same latent construct.

\subsection{O*NET-based own-measure}
The second release is a sparse O*NET-based own-measure. It uses the same 232 standardized O*NET attributes considered in Section 7, but it is constructed differently. The O*NET anchor in Section 7 is unsupervised and is used as a validation device. The own-measure in this section is supervised: it is obtained by regressing the corrected exposure score on the O*NET attributes using LASSO with ten-fold cross-validation on the merged occupation sample. 

At the selected regularization parameter, the LASSO retains 83 features. The selected variables include cognitive and analytical attributes such as working with computers, written comprehension, electronic-mail use, reading comprehension, operations analysis, and management of financial resources. They also include physical or manual attributes such as multilimb coordination, extent flexibility, static strength, finger dexterity, and repairing. This feature pattern is consistent with the routine-versus-analytical dimension emphasized in earlier task-based work and with the principal-component evidence in Section 7. 

The LASSO prediction approximates the corrected score closely. Its Spearman correlation with the cross-source-corrected exposure score is 0.962 on the merged sample. Its Spearman correlation with the Felten et al. index is 0.947. These correlations indicate that much of the corrected score can be reproduced from conventional O*NET attributes, without conditioning directly on large-language-model outputs. 

The own-measure does not replace the multi-source correction. Rather, it is a deployable approximation to the corrected score. The partial-identification band attached to the corrected score is inherited from the MS-GMM construction, not from the LASSO regression itself. Researchers using the own-measure should therefore treat it as an approximation to the corrected exposure construct and should account for any additional approximation error when the distinction is material.

\subsection{Limitations and versioning}
The released scores are occupation-level measures. They do not capture within-occupation heterogeneity in worker-level AI use or task assignment. The partial-identification band therefore applies to the occupation-level exposure construct. It does not bound the additional error that would arise from assigning heterogeneous worker-level exposure intensities within the same occupation. Incorporating worker-level AI-use intensity would require repeated micro-survey measures or other data that identify within-occupation variation over time.

The score table is also static with respect to model vintage. The language-model sources reflect the capabilities and annotation behavior of the specific releases used to construct the panel. As new rater versions, rubrics, and worker-survey waves become available, the measurement functions may change. The release is therefore versioned. Future releases can incorporate new sources while preserving information on the score version used in downstream empirical work

\section{Conclusion}
This paper has studied regression on a latent variable whose available measurements are noisy and possibly nonlinear. No single source can point-identify the structural coefficient, because each source carries its own slope and curvature. The multi-source structure nevertheless delivers informative partial identification: bounding curvature relative to slope confines the coefficient to a closed-form interval around the symmetric cross-source estimator, with width second order in the bound and free of the unknown source loadings. Once four measurements are available the bound itself becomes estimable, through the split-instrument regression that corrects the errors-in-variables attenuation of its direct OLS counterpart, and the Imbens-Manski procedure with the Stoye critical value turns the estimated endpoints into inference that is uniform over the curvature class, including at the point-identified boundary.

The empirical application applies the method to occupational measures of exposure to artificial intelligence. Six candidate measures produce substantially different downstream employment coefficients when merged into the same American Community Survey panel. An ex ante factor-analytic screen separates the Webb patent-text measure from the language-model-and-Felten group, indicating that the patent-text score captures a distinct construct. For the five retained measures, the proposed estimator yields a loading-invariant consensus coefficient of \(- 0.239\). The corresponding partial-identification half-width is 1.23 percent of the point estimate at the estimated curvature bound and 1.88 percent when the bound is replaced by its one-sided 95 percent upper confidence bound. In this application, residual nonlinear-measurement uncertainty is small relative to sampling uncertainty. The main empirical lesson is that construct choice matters more than the remaining curvature bias within the retained construct.

The paper also provides occupation-level outputs for downstream use. The released score table reports a corrected AI-exposure measure for 682 six-digit Standard Occupational Classification codes, together with partial-identification bands. A companion O*NET-based own-measure approximates the corrected score without using large-language-model output. These objects are intended to make the measurement correction transparent and replicable in future empirical work.

Although the application concerns AI exposure, the identification problem is more general. Similar multi-source measurement structures arise in ESG ratings, LLM-based annotation, survey coding, and other settings in which several noisy sources measure the same latent object through possibly different transformations. The results apply when the analyst has at least four measurements, can assess the cross-source orthogonality restrictions, and can estimate the curvature bound using the split-instrument diagnostic.

\clearpage
\section*{References}
\addcontentsline{toc}{section}{References}

Acemoglu, Daron. 2025. ``The Simple Macroeconomics of AI.'' \emph{Economic Policy} 40 (121): 13-58.

Acemoglu, Daron, and David Autor. 2011. ``Skills, Tasks and Technologies: Implications for Employment and Earnings.'' In \emph{Handbook of Labor Economics}, Vol. 4B, edited by Orley Ashenfelter and David Card, 1043-1171. Amsterdam: Elsevier.

Acemoglu, Daron, and Pascual Restrepo. 2022. ``Tasks, Automation, and the Rise in U.S. Wage Inequality.'' \emph{Econometrica} 90 (5): 1973-2016. \url{https://doi.org/10.3982/ECTA19815}.

Allman, Elizabeth S., Catherine Matias, and John A. Rhodes. 2009. ``Identifiability of Parameters in Latent Structure Models with Many Observed Variables.'' \emph{Annals of Statistics} 37 (6A): 3099-3132. \url{https://doi.org/10.1214/09-AOS689}.

Andrews, Isaiah, Matthew Gentzkow, and Jesse M. Shapiro. 2017. ``Measuring the Sensitivity of Parameter Estimates to Estimation Moments.'' \emph{Quarterly Journal of Economics} 132 (4): 1553-1592.

Appel, Ruth, Peter McCrory, Alex Tamkin, Miles McCain, Tyler Neylon, and Michael Stern. 2025. ``Anthropic Economic Index Report: Uneven Geographic and Enterprise AI Adoption.'' arXiv:2511.15080. \url{https://arxiv.org/abs/2511.15080}.

Armstrong, Timothy B., and Michal Kolesár. 2018. ``Optimal Inference in a Class of Regression Models.'' \emph{Econometrica} 86 (2): 655-683.

Armstrong, Timothy B., and Michal Kolesár. 2020. ``Simple and Honest Confidence Intervals in Nonparametric Regression.'' \emph{Quantitative Economics} 11 (1): 1-39.

Autor, David H., Frank Levy, and Richard J. Murnane. 2003. ``The Skill Content of Recent Technological Change: An Empirical Exploration.'' \emph{Quarterly Journal of Economics} 118 (4): 1279-1333. \url{https://doi.org/10.1162/003355303322552801}.

Berg, Florian, Julian F. Kölbel, and Roberto Rigobon. 2022. ``Aggregate Confusion: The Divergence of ESG Ratings.'' \emph{Review of Finance} 26 (6): 1315-1344. \url{https://doi.org/10.1093/rof/rfac033}.

Bertrand, Marianne, Esther Duflo, and Sendhil Mullainathan. 2004. ``How Much Should We Trust Differences-in-Differences Estimates?'' \emph{Quarterly Journal of Economics} 119 (1): 249-275.

Bick, Alexander, Adam Blandin, and David J. Deming. 2024. ``The Rapid Adoption of Generative AI.'' NBER Working Paper 32966. \url{https://www.nber.org/papers/w32966}.

Bick, Alexander, Adam Blandin, David J. Deming, and Tyler Schumacher. 2026. ``What Work Does Generative AI Do?'' Federal Reserve Bank of St.~Louis Working Paper, April 27, 2026.

Bound, John, Charles Brown, and Nancy Mathiowetz. 2001. ``Measurement Error in Survey Data.'' In \emph{Handbook of Econometrics}, Vol. 5, edited by James J. Heckman and Edward Leamer, 3705-3843. Amsterdam: Elsevier.

Brynjolfsson, Erik, Bharat Chandar, and Ruyu Chen. 2025. ``Canaries in the Coal Mine? Six Facts about the Recent Employment Effects of Artificial Intelligence.'' Stanford Digital Economy Lab Working Paper.

Callaway, Brantly, and Pedro H. C. Sant'Anna. 2021. ``Difference-in-Differences with Multiple Time Periods.'' \emph{Journal of Econometrics} 225 (2): 200-230. \url{https://doi.org/10.1016/j.jeconom.2020.12.001}.

Cameron, A. Colin, Jonah B. Gelbach, and Douglas L. Miller. 2011. ``Robust Inference With Multiway Clustering.'' \emph{Journal of Business and Economic Statistics} 29 (2): 238-249.

Cameron, A. Colin, and Douglas L. Miller. 2015. ``A Practitioner's Guide to Cluster-Robust Inference.'' \emph{Journal of Human Resources} 50 (2): 317-372.

Canay, Ivan A., and Azeem M. Shaikh. 2017. ``Practical and Theoretical Advances in Inference for Partially Identified Models. In \emph{Advances in Economics and Econometrics}, Vol. 2, Cambridge: Cambridge University Press.

Chatterji, Aaron, Thomas Cunningham, David J. Deming, Zoe Hitzig, Christopher Ong, Carl Yan Shan, and Kevin Wadman. 2025. ``How People Use ChatGPT.'' NBER Working Paper 34255. \url{https://www.nber.org/papers/w34255}.

Chen, Danqing, Carina Kane, Austin Kozlowski, Nadav Kunievsky, and James A. Evans. 2025. ``The (Short-Term) Effects of Large Language Models on Unemployment and Earnings.'' arXiv:2509.15510. \url{https://arxiv.org/abs/2509.15510}.

Chen, Xiaohong, Timothy M. Christensen, and Elie Tamer. 2018. ``Monte Carlo Confidence Sets for Identified Sets.'' \emph{Econometrica} 86 (6): 1965-2018.

Dawid, A. Philip, and Allan M. Skene. 1979. ``Maximum Likelihood Estimation of Observer Error-Rates Using the EM Algorithm.'' \emph{Journal of the Royal Statistical Society Series C} 28 (1): 20-28. \url{https://www.jstor.org/stable/2346806}.

Chernozhukov, Victor, Denis Chetverikov, Mert Demirer, Esther Duflo, Christian Hansen, Whitney Newey, and James Robins. 2018. ``Double/Debiased Machine Learning for Treatment and Structural Parameters.'' \emph{Econometrics Journal} 21 (1): C1-C68. \url{https://doi.org/10.1111/ectj.12097}.

Eloundou, Tyna, Sam Manning, Pamela Mishkin, and Daniel Rock. 2024. ``GPTs are GPTs: Labor Market Impact Potential of LLMs.'' \emph{Science} 384 (6702): 1306-1308. \url{https://doi.org/10.1126/science.adj0998}.

Fang, Zheng, and Andres Santos. 2019. ``Inference on Directionally Differentiable Functions.'' \emph{Review of Economic Studies} 86 (1): 377-412.

Felten, Edward W., Manav Raj, and Robert Seamans. 2021. ``Occupational, Industry, and Geographic Exposure to Artificial Intelligence: A Novel Dataset and Its Potential Uses.'' \emph{Strategic Management Journal} 42 (12): 2195-2217. \url{https://doi.org/10.1002/smj.3286}.

Frey, Carl Benedikt, and Michael A. Osborne. 2017. ``The Future of Employment: How Susceptible are Jobs to Computerisation?'' \emph{Technological Forecasting and Social Change} 114: 254-280.

Gillen, Ben, Erik Snowberg, and Leeat Yariv. 2019. ``Experimenting with Measurement Error: Techniques with Applications to the Caltech Cohort Study.'' \emph{Journal of Political Economy} 127 (4): 1826-1863. \url{https://doi.org/10.1086/701808}.

Goodman, Leo A. 1974. ``Exploratory Latent Structure Analysis Using Both Identifiable and Unidentifiable Models.'' \emph{Biometrika} 61 (2): 215-231.

Hampole, Menaka, Dimitris Papanikolaou, Lawrence D.W. Schmidt, and Bryan Seegmiller. 2025. ``Artificial Intelligence and the Labor Market.'' NBER Working Paper 33509. \url{https://www.nber.org/papers/w33509}.

Handa, Kunal, Alex Tamkin, Miles McCain, Saffron Huang, Esin Durmus, Sarah Heck, Jared Mueller, et al. 2025. ``Which Economic Tasks are Performed with AI? Evidence from Millions of Claude Conversations.'' Anthropic Technical Report.

Hansen, Lars Peter. 1982. ``Large Sample Properties of Generalized Method of Moments Estimators.'' \emph{Econometrica} 50 (4): 1029-1054.

Horn, John L. 1965. ``A Rationale and Test for the Number of Factors in Factor Analysis.'' \emph{Psychometrika} 30 (2): 179-185.

Hu, Yingyao, and Susanne M. Schennach. 2008. ``Instrumental Variable Treatment of Nonclassical Measurement Error Models.'' \emph{Econometrica} 76 (1): 195-216.

Humlum, Anders, and Emilie Vestergaard. 2025. ``Still Waters, Rapid Currents: Early Labor Market Transformation under Generative AI.'' NBER Working Paper 33777. \url{https://www.nber.org/papers/w33777}.

Imbens, Guido W., and Charles F. Manski. 2004. ``Confidence Intervals for Partially Identified Parameters.'' \emph{Econometrica} 72 (6): 1845-1857. \url{https://doi.org/10.1111/j.1468-0262.2004.00555.x}.

Kaido, Hiroaki, Francesca Molinari, and Jórg Stoye. 2019. ``Confidence Intervals for Projections of Partially Identified Parameters.'' \emph{Econometrica} 87 (4): 1397-1432.

Lazarsfeld, Paul F. 1950. ``The Logical and Mathematical Foundation of Latent Structure Analysis.'' In \emph{Measurement and Prediction}, edited by S.~A.~Stouffer et al., 362-412. Princeton: Princeton University Press.

Lee, David S. 2009. ``Training, Wages, and Sample Selection: Estimating Sharp Bounds on Treatment Effects.'' \emph{Review of Economic Studies} 76 (3): 1071-1102.

Liang, Kung-Yee, and Scott L. Zeger. 1986. ``Longitudinal Data Analysis Using Generalized Linear Models.'' \emph{Biometrika} 73 (1): 13-22.

Machovec, Christine, Michael J. Rieley, and Emily Rolen. 2025. ``Incorporating AI Impacts in BLS Employment Projections: Occupational Case Studies.'' \emph{Monthly Labor Review}, U.S. Bureau of Labor Statistics, February.

Manski, Charles F. 2003. \emph{Partial Identification of Probability Distributions}. New York: Springer.

Massenkoff, Maxim, and Peter McCrory. 2026. ``Labor Market Impacts of AI: A New Measure and Early Evidence.'' Anthropic Research, March 5.

Molinari, Francesca. 2020. ``Microeconometrics with Partial Identification. In \emph{Handbook of Econometrics}, Vol. 7A, edited by Steven N. Durlauf, Lars Peter Hansen, James J. Heckman, and Rosa L. Matzkin, 355-486. Amsterdam: Elsevier.

Ogut, Burhan, and Michelle Yin. 2026. ``Identification and Inference with Multiple LLM Annotators: Correcting Non-Classical Measurement Error in AI Exposure.'' RISEI Lab Working Paper, Northwestern University.

Rambachan, Ashesh, and Jonathan Roth. 2023. ``A More Credible Approach to Parallel Trends.'' \emph{Review of Economic Studies} 90 (5): 2555-2591. \url{https://doi.org/10.1093/restud/rdad018}.

Schennach, Susanne M. 2016. ``Recent Advances in the Measurement Error Literature.'' \emph{Annual Review of Economics} 8: 341-377.

Stock, James H., and Motohiro Yogo. 2005. ``Testing for Weak Instruments in Linear IV Regression.'' In \emph{Identification and Inference for Econometric Models: Essays in Honor of Thomas Rothenberg}, edited by Donald W. K. Andrews and James H. Stock, 80-108. Cambridge: Cambridge University Press.

Stoye, Jörg. 2009. ``More on Confidence Intervals for Partially Identified Parameters.'' \emph{Econometrica} 77 (4): 1299-1315. \url{https://doi.org/10.3982/ECTA7347}.

Sun, Liyang, and Sarah Abraham. 2021. ``Estimating Dynamic Treatment Effects in Event Studies with Heterogeneous Treatment Effects.'' \emph{Journal of Econometrics} 225 (2): 175-199. \url{https://doi.org/10.1016/j.jeconom.2020.09.006}.

Tamer, Elie. 2010. ``Partial Identification in Econometrics.'' \emph{Annual Review of Economics} 2: 167-195.

Tomlinson, Kiran, Sonia Jaffe, Will Wang, Scott Counts, and Siddharth Suri. 2025. ``Working with AI: Measuring the Applicability of Generative AI to Occupations.'' Microsoft Research Technical Report (arXiv:2507.07935).

Webb, Michael. 2020. ``The Impact of Artificial Intelligence on the Labor Market.'' SSRN Working Paper 3482150.

Yin, Michelle, and Burhan Ogut. 2026. ``Platform Selection in AI-Exposure Measurement: Who Uses AI and Why It Matters for Downstream Estimates.'' Working Paper, Northwestern University.

Yin, Michelle, Hoa Vu, and Claudia Persico. 2026. ``How (un)Stable Are LLM Occupational Exposure Scores? Evidence from Multi-Model Replication.'' NBER Working Paper 35110.

\renewcommand{\thesection}{Appendix \Alph{section}.}
\renewcommand{\theequation}{\Alph{section}.\arabic{equation}}
\setcounter{section}{0}
\setcounter{equation}{0}

\renewcommand{\thesubsection}{\Alph{section}.\arabic{subsection}}

\section{Proofs}

This appendix proves Proposition 1, Theorem 1, Corollaries 1 and 2, the split-instrument curvature lemma, Proposition 2, and Theorem 2. The notation follows Sections 2 through 5. Observed measurements are denoted by \(M_{i}^{(s)}\). All variables are understood as Frisch-Waugh-Lovell residuals with respect to \(X_{i}\) unless stated otherwise.

For each source \(s\),

\[M_{i}^{(s)} = \mu_{s}(E_{i}) + U_{i}^{(s)},\]

and, by Assumptions 3 and 4,

\begin{equation}
\mu_s(E) = a_s + c_s E + \tfrac{1}{2}\gamma_s E^2 + R_s(E).
\end{equation}

The Lagrange remainder satisfies

\[R_{s}(E) = \frac{1}{6}\mu_{s}^{'''}(\xi_{s}(E))E^{3}\]

for some \(\xi_{s}(E)\) between 0 and \(E\), and therefore

\[\lvert R_{s}(E) \rvert \leq \frac{1}{6}CB^{2}c_{s} \lvert E \rvert^{3}.\]

Assumption 5 implies \(\mathbb{E}[ R_{s}(E_{i})^{2}] < \infty\). For a partition of the source set into \(S_{1}\) and \(S_{2}\), define

\begin{equation}
\bar M_i^{(j)} = \bar a_j + \bar c_j E_i + \tfrac{1}{2}\bar\gamma_j E_i^2 + \bar U_i^{(j)} + \bar R_i^{(j)}, \qquad j = 1, 2,
\end{equation}

where \({\bar{a}}_{j}\), \({\bar{c}}_{j}\), \({\bar{\gamma}}_{j}\), \({\bar{U}}_{i}^{(j)}\), and \({\bar{R}}_{i}^{(j)}\) are subset averages. Let

\[h_{j} = \frac{{\bar{\gamma}}_{j}}{{\bar{c}}_{j}}.\]

Since \(c_{s} > 0\) and \(\lvert \gamma_{s} \rvert \leq Bc_{s}\), \(\lvert h_{j} \rvert \leq B\).

\subsection{Proof of Proposition 1}
Let

\[\sigma_{Yj} = \operatorname{Cov}(Y_{i},{\bar{M}}_{i}^{(j)}), \qquad \sigma_{12} = \operatorname{Cov}({\bar{M}}_{i}^{(1)},{\bar{M}}_{i}^{(2)}).\]

Using the structural equation~\eqref{eq:struct} and \(\mathbb{E}[\varepsilon_{i} \mid E_{i},X_{i}] = 0\),

\[\sigma_{Yj} = \beta{\bar{c}}_{j}\operatorname{Var}(E_{i}) + \frac{1}{2}\beta{\bar{\gamma}}_{j}\operatorname{Cov}(E_{i},E_{i}^{2}) + \beta \operatorname{Cov}(E_{i},{\bar{R}}_{i}^{(j)}).\]

The covariance terms involving \({\bar{U}}_{i}^{(j)}\) vanish by Assumption 2, and the covariance terms involving \(\varepsilon_{i}\) vanish by the structural exogeneity condition. Under Assumption 1,

\[\operatorname{Var}(E_{i}) = 1, \qquad \operatorname{Cov}(E_{i},E_{i}^{2}) = \kappa_{3}.\]

Define

\[\rho_{j} = \operatorname{Cov}(E_{i},{\bar{R}}_{i}^{(j)}).\]

By Cauchy-Schwarz and the bound on the Taylor remainder,

\[\rho_{j} = O(B^{2}{\bar{c}}_{j}).\]

Hence

\begin{equation}
\sigma_{Yj} = \beta{\bar{c}}_{j}\left( 1+\frac{1}{2}h_{j}\kappa_{3} \right) + \beta\rho_{j} + O(B^{3}).
\end{equation}

Next, expanding the denominator covariance and applying Assumption 2 to cross-source noise terms gives

\[\sigma_{12} = {\bar{c}}_{1}{\bar{c}}_{2} + \frac{1}{2}{\bar{c}}_{1}{\bar{\gamma}}_{2}\kappa_{3} + \frac{1}{2}{\bar{c}}_{2}{\bar{\gamma}}_{1}\kappa_{3} + \frac{1}{4}{\bar{\gamma}}_{1}{\bar{\gamma}}_{2}(\kappa_{4} - 1) + {\bar{c}}_{1}\rho_{2} + {\bar{c}}_{2}\rho_{1} + O(B^{3}).\]

Equivalently,

\begin{equation}
\sigma_{12} = {\bar{c}}_{1}{\bar{c}}_{2}D + P + O(B^{3}),
\end{equation}

where

\[D = 1 + \frac{1}{2}(h_{1} + h_{2})\kappa_{3} + \frac{1}{4}h_{1}h_{2}(\kappa_{4} - 1),\]

and

\[P = {\bar{c}}_{1}\rho_{2} + {\bar{c}}_{2}\rho_{1}.\]

The condition \(D > 0\) ensures that the population ratios are well defined.

The asymmetric two-stage least-squares estimand is

\begin{equation}
\beta_{2SLS} = \frac{\sigma_{Y2}}{\sigma_{12}},
\end{equation}

A first-order expansion of the ratio gives

\[\operatorname{plim}\beta_{2SLS} = \frac{\beta}{{\bar{c}}_{1}}\frac{1 + \frac{1}{2}h_{2}\kappa_{3}}{D} + O(B^{2}).\]

The \(O(B^{2})\) remainder includes the uncancelled Taylor-remainder covariance. Thus the asymmetric ratio is not invariant to the loading of the endogenous measurement and, even under homogeneous loadings, contains first-order skewness bias when \(\kappa_{3} \neq 0\).

The symmetric estimand is

\begin{equation}
\beta_{sym} = \operatorname{sign}\!\left(\frac{\sigma_{Y2}}{\sigma_{12}}\right)\left(\frac{\sigma_{Y1}\,\sigma_{Y2}}{\sigma_{12}}\right)^{1/2},
\end{equation}

Let

\[A_{j} = 1 + \frac{1}{2}h_{j}\kappa_{3}, \qquad Q = {\bar{c}}_{1}{\bar{c}}_{2}.\]

Then

\[\sigma_{Y1}\sigma_{Y2} = \beta^{2}[ QA_{1}A_{2} + P] + O(B^{3}),\]

because \(A_{j} = 1 + O(B)\) and \(\rho_{j} = O(B^{2})\). Since

\[\sigma_{12} = QD + P + O(B^{3}),\]

the common term \(P\) cancels in the ratio to the required order, and

\[\frac{\sigma_{Y1}\sigma_{Y2}}{\sigma_{12}} = \beta^{2}\frac{A_{1}A_{2}}{D} + O(B^{3}).\]

Since

\[A_{1}A_{2} = 1 + \frac{1}{2}(h_{1} + h_{2})\kappa_{3} + \frac{1}{4}h_{1}h_{2}\kappa_{3}^{2},\]

we obtain

\[(\operatorname{plim}\beta_{sym})^{2} = \beta^{2}\frac{1 + \frac{1}{2}(h_{1} + h_{2})\kappa_{3} + \frac{1}{4}h_{1}h_{2}\kappa_{3}^{2}}{D} + O(B^{3}).\]

The sign convention gives

\[\operatorname{sign}(\operatorname{plim}\beta_{sym}) = \operatorname{sign}(\beta)\]

for \(B\) sufficiently small. This proves Proposition 1.

\subsection{Proof of Theorem 1}
Let

\[N = 1 + \frac{1}{2}(h_{1} + h_{2})\kappa_{3} + \frac{1}{4}h_{1}h_{2}\kappa_{3}^{2}.\]

By Proposition 1,

\begin{equation}
\frac{\left( \operatorname{plim}\beta_{sym} \right)^{2}}{\beta^{2}} = \frac{N}{D} + O(B^{3}).
\end{equation}

Moreover,

\[D - N = \frac{1}{4}h_{1}h_{2}(\kappa_{4} - 1 - \kappa_{3}^{2}).\]

The term \(\kappa_{4} - 1 - \kappa_{3}^{2}\) is nonnegative. To see this, project \(E_{i}^{2}\) on \(\left( 1,E_{i} \right)\). Under Assumption 1, the residual is

\[E_{i}^{2} - 1 - \kappa_{3}E_{i},\]

and its variance is

\[\operatorname{Var}(E_{i}^{2}) - \frac{\operatorname{Cov}(E_{i},E_{i}^{2})^{2}}{\operatorname{Var}(E_{i})} = \kappa_{4} - 1 - \kappa_{3}^{2} \geq 0.\]

For \(B\) small enough that \(D\) is bounded away from zero,

\[\frac{N}{D} - 1 = - \frac{h_{1}h_{2}}{4D}(\kappa_{4} - 1 - \kappa_{3}^{2}) = O(B^{2}).\]

Taking square roots around one gives

\[\sqrt{\frac{N}{D}} - 1 = - \frac{h_{1}h_{2}}{8D}(\kappa_{4} - 1 - \kappa_{3}^{2}) + O(B^{3}).\]

Since \(\lvert h_{j} \rvert \leq B\) and \(D = 1 + O(B)\),

\begin{equation}
\lvert \sqrt{\frac{N}{D}} - 1 \rvert \leq \frac{1}{8}B^{2}(\kappa_{4} - 1 - \kappa_{3}^{2}) + O(B^{3}).
\end{equation}

Thus

\begin{equation}
\lvert \beta - \operatorname{plim}\beta_{sym} \rvert \leq \lvert \operatorname{plim}\beta_{sym} \rvert \Delta + O(B^{3}),
\end{equation}

where

\begin{equation}
\Delta = \tfrac{1}{8}B^2(\kappa_4 - 1 - \kappa_3^2).
\end{equation}

The loadings do not enter this bound because they cancel in the symmetric covariance product.

Sharpness follows from the second-order expansion. The second-order deviation is proportional to \(- h_{1}h_{2}\). The endpoints are attained, up to \(O(B^{3})\), at same-sign and opposite-sign corner configurations of the box \([ - B,B]^{2}\). In particular, \(h_{1} = h_{2} = B\) and \(h_{1} = - h_{2} = B\) attain the two endpoint deviations after taking the square root. Continuity of the map

\[(h_{1},h_{2}) \mapsto \sqrt{N/D} - 1\]

on the compact box implies that every intermediate value in the interval is attained by some admissible curvature pair. Such curvature pairs are feasible by choosing source-specific relative curvatures constant within each subset. This proves Theorem 1.

\subsection{Proof of Corollary 1}
If \(c_{s} = c\) for all \(s\), then

\[{\bar{c}}_{1} = {\bar{c}}_{2} = c.\]

From the asymmetric expansion in Proposition 1,

\[c\text{ }\operatorname{plim}\beta_{2SLS} = \beta\frac{1 + \frac{1}{2}h_{2}\kappa_{3}}{D} + O(B^{2}).\]

Let

\[N_{a} = 1 + \frac{1}{2}h_{2}\kappa_{3}.\]

Then

\[N_{a} - D = - \frac{1}{2}h_{1}\kappa_{3} - \frac{1}{4}h_{1}h_{2}(\kappa_{4} - 1).\]

Therefore, uniformly over \(\lvert h_{j} \rvert \leq B\),

\[\lvert \frac{N_{a}}{D} - 1 \rvert \leq \frac{1}{2}B \lvert \kappa_{3} \rvert + \frac{1}{4}B^{2} \lvert \kappa_{4} - 1 \rvert + O(B^{2}),\]

where the displayed \(O(B^{2})\) also includes the uncancelled Taylor-remainder covariance in the asymmetric ratio. This gives the stated asymmetric half-width, with an overall \(O(B^{2})\) approximation remainder. The leading term is first order in \(B\) whenever the latent distribution is skewed.

\subsection{Proof of Corollary 2}
If \(E_{i}\) is standard normal, then

\[\kappa_{3} = 0, \qquad \kappa_{4} = 3.\]

Substituting these values into Theorem 1 yields

\[\Delta = \frac{1}{8}B^{2}(3 - 1) = \frac{B^{2}}{4},\]

up to the stated higher-order remainder.

\subsection{Split-instrument curvature lemma}
Fix a source \(s\) and partition the remaining sources into two disjoint sets \(A_{s}\) and \(B_{s}\), with leave-out averages \(Z_{A}\) and \(Z_{B}\). Up to the Taylor remainders of Assumption 3,
\[M^{(s)} = a_{s} + c_{s}E + \tfrac{1}{2}\gamma_{s}E^{2} + U^{(s)}, \qquad Z_{A} = a_{A} + {\bar{c}}_{A}E + \tfrac{1}{2}{\bar{\gamma}}_{A}E^{2} + U_{A},\]
and similarly for \(Z_{B}\) with loading \({\bar{c}}_{B}\) and curvature \({\bar{\gamma}}_{B}\). The auxiliary equation regresses \(M^{(s)}\) on \((1, Z_{A}, Z_{A}^{2})\) with instruments \((1, Z_{B}, Z_{B}^{2})\). Under Assumption 2 the noise terms are mutually mean-independent given \(E\) and have conditional variances that do not depend on \(E\), so the noise in \(Z_{A}\) enters the instrumented moment equations only through terms that vanish or cancel after centering.

Write \(g = \tfrac{1}{2}\gamma_{s}\), \(q = \tfrac{1}{2}{\bar{\gamma}}_{A}\), \(a = {\bar{c}}_{A}\), and \(b_{1} = c_{s}/a + \delta\). Evaluating the two instrumented moment equations at the population parameters and retaining every term of first order in the curvatures gives
\[a\,\delta + a^{2}\kappa_{3}\,b_{2} = \kappa_{3}\left( g - q\,c_{s}/a \right) + O(B^{2}),\]
\[a\,\kappa_{3}\,\delta + a^{2}(\kappa_{4}-1)\,b_{2} = (\kappa_{4}-1)\left( g - q\,c_{s}/a \right) + O(B^{2}).\]
The terms involving the instrument curvature \({\bar{\gamma}}_{B}\) cancel between the two sides of the second equation. Solving the system,
\[b_{2} = \frac{g - q\,c_{s}/a}{a^{2}} + O(B^{2}), \qquad b_{1} = \frac{c_{s}}{a} + O(B^{2}).\]
Since \(\operatorname{Cov}(Z_{A},Z_{B}) = {\bar{c}}_{A}{\bar{c}}_{B} + O(B)\),
\[g_{s} = 2 b_{2}\,\frac{\left\{ \operatorname{Cov}(Z_{A},Z_{B}) \right\}^{1/2}}{b_{1}} = \left( \frac{\gamma_{s}}{c_{s}} - \frac{{\bar{\gamma}}_{A}}{{\bar{c}}_{A}} \right)\left( \frac{{\bar{c}}_{B}}{{\bar{c}}_{A}} \right)^{1/2} + O(B^{2}).\]
The estimand is the curvature of source \(s\) relative to its leave-out benchmark, scaled by the square root of the loading ratio. Three consequences follow. First, at the linear benchmark \(\gamma_{s} = {\bar{\gamma}}_{A} = {\bar{\gamma}}_{B} = 0\) the probability limit is zero, so the diagnostic generates no spurious curvature. Second, a curvature component common to every source drops out of the contrast; such a component is absorbed by the normalization \(\sum_{s}\gamma_{s} = 0\) of Section 2 and is not identifiable from the joint distribution of the sources. Third, the scale factor equals one when the leave-out loadings coincide, so the relevant condition on the split concerns the loadings rather than the number of sources in each subset; after per-source standardization the loadings implied by the observed cross-source correlations differ by a few percent, and the factor is close to one.

Under the normalization the contrasts identify the individual departures. With the leave-out designs used in the application, the population contrasts and the restriction \(\sum_{s}\gamma_{s}/c_{s} = 0\) determine the departures \(\gamma_{s}/c_{s}\), and the estimated bound
\[\widehat{B} = \max_{s}\,\lvert {\widehat{g}}_{s} \rvert\]
converges at the parametric rate to the maximal contrast under Assumption 5. By Assumption 6 the maximizing source is unique, so the limit distribution of \(\widehat{B}\) follows from the delta method; at a tie the maximum is directionally differentiable, as in Fang and Santos (2019). In the empirical application the subset-average curvatures implied by the contrasts are well below the maximal contrast, so evaluating the half-width at \(\widehat{B}\) is conservative for the configuration the data indicate.

\subsection{Proof of Proposition 2}
Let \(\widehat{m}\) denote the moment vector defined in Section 5.1. Its components consist of sample covariances, IV auxiliary-regression moments, and standardized moments of the cross-source average. Under Assumption 5, these components are sample averages, or smooth functions of sample averages, with sufficient finite moments. Therefore,

\begin{equation}
\sqrt{n}(\widehat m - m) \xrightarrow{d} \mathcal{N}(0, \Omega),
\end{equation}

Under clustered sampling, \(\Omega\) is replaced by the corresponding cluster-robust covariance matrix.

The endpoint map is differentiable at the population point under the maintained regularity conditions. The symmetric estimator is differentiable when \(\sigma_{12}\) is bounded away from zero and the radicand is positive. The curvature bound is differentiable at a unique maximizer by Assumption 6. The standardized moments are smooth functions of the underlying sample moments. The estimated half-width is a smooth composition of these quantities. Finally, the lower and upper endpoints are oriented by minimum and maximum; away from the zero-width crossing and under the endpoint-variance condition, this orientation is locally differentiable. The delta method gives

\[\begin{aligned}
\sqrt{n}\left[ \left( \begin{array}{r}
{\widehat{\theta}}_{l} \\
{\widehat{\theta}}_{u}
\end{array} \right)-\left( \begin{array}{r}
\theta_{l} \\
\theta_{u}
\end{array} \right) \right] \xrightarrow{d} N(0,\Sigma_{\theta}),
\end{aligned}\]

where

\[\Sigma_{\theta} = J(m)\Omega J(m)'.\]

This proves Proposition 2.

\subsection{Proof of Theorem 2}
The confidence interval in Section 5 is the Imbens-Manski interval with the Stoye critical value. We verify the required conditions uniformly over the class \(\mathcal{P}(B_{\max})\).

First, Proposition 2 gives joint asymptotic normality of the estimated endpoints. The bounded eighth-moment condition implies the Lindeberg condition uniformly over \(\mathcal{P}(B_{\max})\), and the restrictions that \(\sigma_{12}\) and endpoint variances are bounded away from zero rule out singular limiting cases.

Second, the plug-in covariance estimator is uniformly consistent. This follows from the same bounded-moment restrictions, the continuity of the endpoint map, and standard covariance consistency for the relevant independent or clustered sampling scheme.

Third, the estimated width

\[{\widehat{\theta}}_{u} - {\widehat{\theta}}_{l}\]

is root-\(n\) consistent for the population width. The point-identified boundary corresponds to zero population width. The Stoye critical value is designed for this boundary case and for sequences approaching it.

Therefore Stoye\textquotesingle s theorem implies

\[\underset{n}{\liminf}\underset{P \in \mathcal{P}(B_{n})}{\inf}P_{P}\{\beta \in C_{n}(1 - \alpha)\} \geq 1 - \alpha\]

for \(\beta\) in the population identified set. Since Theorem 1 characterizes that set up to a third-order approximation error, the coverage statement for the original structural parameter holds up to that approximation. The location error of the estimated endpoints induced by the curvature approximation is of order \(B_{n}^{3}\), while the sampling margins of the interval are of order \(n^{-1/2}\). Under the maintained rate condition \(\sqrt{n}\,B_{n}^{3} \rightarrow 0\) the induced coverage error vanishes. This proves Theorem 2.

~

\section{Platform-Selection Extension}
\setcounter{equation}{0}

The main text treats sources as indexed only by \(s\). In some applications, however, a source may also be observed through a platform-specific sampling process. In the AI-exposure setting, platform conversation shares may differ from occupational workforce shares. This creates a potential platform-selection component in the measured exposure score.

Let \(p = 1,\ldots,P\) index platforms. Suppose the observed source-platform measurement satisfies

\begin{equation}
M_{i}^{\left( s,p \right)} = \mu_{s}(E_{i}) + \lambda_{p}\psi_{o(i),p} + U_{i}^{\left( s,p \right)},
\end{equation}

where \(\psi_{o,p}\) is the platform-\(p\) selection ratio for occupation \(o\), defined as the ratio of platform \(p\)\textquotesingle s conversation share for occupation \(o\) to that occupation\textquotesingle s workforce share. The scalar \(\lambda_{p}\) measures how platform selection loads into the observed measurement. The disturbance \(U_{i}^{\left( s,p \right)}\) is source-platform measurement error.

A direct regression of \(M_{i}^{\left( s,p \right)}\) on \(\psi_{o(i),p}\) would identify \(\lambda_{p}\) only under an orthogonality condition between platform selection and latent exposure. That condition is not innocuous in this application, since occupations with high platform usage may also differ systematically in latent AI applicability. We therefore use within-source cross-platform differences. The differencing removes the common measurement component \(\mu_{s}(E_{i})\) and identifies the platform-selection loadings from cross-platform variation in the selection ratios.

\textbf{Lemma B.1. Within-source cross-platform identification.}\\
Fix a reference source \(s_{0}\). For each pair of platforms \(p \neq p'\), define the within-source difference

\[D_{i}^{\left( p,p' \right)} = M_{i}^{\left( s_{0},p \right)} - M_{i}^{\left( s_{0},p' \right)}.\]

Then

\begin{equation}
D_{i}^{\left( p,p' \right)} = \lambda_{p}\psi_{o(i),p} - \lambda_{p'}\psi_{o(i),p'} + U_{i}^{\left( s_{0},p \right)} - U_{i}^{\left( s_{0},p' \right)}.
\end{equation}

Stack the \(P(P - 1)/2\) pairwise differences across platform pairs and occupations. If the resulting cross-platform selection-ratio design matrix has full column rank after the within-pair transformation, then \(\left( \lambda_{1},\ldots,\lambda_{P} \right)\) is point identified. Under the moment and sampling conditions in Assumption 5,

\begin{equation}
\sqrt{n}(\widehat{\lambda} - \lambda) \xrightarrow{d} N(0,\Sigma_{\lambda}).
\end{equation}

\textbf{Proof.}\\
For a fixed source \(s_{0}\), the term \(\mu_{s_{0}}(E_{i})\) enters \(M_{i}^{\left( s_{0},p \right)}\) and \(M_{i}^{\left( s_{0},p' \right)}\) identically. It therefore cancels in the within-source cross-platform difference. The remaining equation is linear in the platform-selection loadings. The full-rank condition on the stacked selection-ratio design matrix gives point identification. The composite error in the pairwise-difference equation is \(U_{i}^{\left( s_{0},p \right)} - U_{i}^{\left( s_{0},p' \right)}\). Under the maintained mean-independence restrictions, this error has conditional mean zero given the latent exposure and the selection ratios. Standard least-squares asymptotics then give the stated limiting distribution. \(\square\)

After estimating the platform loadings, define the adjusted measurement

\[{\widetilde{M}}_{i}^{\left( s,p \right)} = M_{i}^{\left( s,p \right)} - {\widehat{\lambda}}_{p}\psi_{o(i),p}.\]

Uniformly over observations with bounded selection ratios,

\[{\widetilde{M}}_{i}^{\left( s,p \right)} = \mu_{s}(E_{i}) + U_{i}^{\left( s,p \right)} + O_{p}(n^{- 1/2}).\]

Averaging \({\widetilde{M}}_{i}^{\left( s,p \right)}\) across platforms within source \(s\) gives the adjusted source-level measurement used in the half-average construction of Section 2. The estimation error in \({\widehat{\lambda}}_{p}\) enters the adjusted measurement at the usual parametric rate and can be incorporated into the moment vector used for inference if the platform correction is material.

This extension identifies and removes the platform-selection component without imposing orthogonality between platform selection and latent exposure. The identifying variation comes from within-source differences across platforms rather than from cross-occupation variation in a single platform.

If platform conversation-share data are unavailable, the analyst cannot separately identify the platform-selection component. In that case, the baseline multi-source estimator may still be applied to the unadjusted measurements, but the platform component is then part of the source-specific measurement error. The validity of the baseline procedure depends on whether this composite error satisfies the cross-source orthogonality restrictions in Assumption 2. If platform selection is likely to induce common residual dependence across sources, the resulting estimates should be interpreted as conditional on that maintained restriction or reported as a sensitivity analysis.

\clearpage
\section*{Tables}
\addcontentsline{toc}{section}{Tables}

\begin{table}[h!]
\centering
\caption{Cross-source correlation matrix}\label{tab:t1}
\small
\begin{tabular}{lcccccc}
\toprule
Source & ChatGPT-5 & Claude 4.5 & GPT-4 & Gemini 2.5 & Felten & Webb \\
\midrule
ChatGPT-5 & 1.000 & 0.901 & 0.915 & 0.929 & 0.851 & 0.079 \\
Claude 4.5 & 0.901 & 1.000 & 0.918 & 0.881 & 0.875 & 0.061 \\
GPT-4 & 0.915 & 0.918 & 1.000 & 0.884 & 0.869 & 0.101 \\
Gemini 2.5 & 0.929 & 0.881 & 0.884 & 1.000 & 0.823 & 0.060 \\
Felten & 0.851 & 0.875 & 0.869 & 0.823 & 1.000 & 0.025 \\
Webb & 0.079 & 0.061 & 0.101 & 0.060 & 0.025 & 1.000 \\
\bottomrule
\end{tabular}
\\[6pt]
\begin{minipage}{0.95\textwidth}
\footnotesize\textbf{Notes.} Pearson correlations among the six AI-exposure measures on the 682-occupation six-digit SOC panel. The four language-model raters and the Felten (2021) ability-based measure are mutually correlated above $0.82$ in pairwise comparison, while the patent-text exposure of Webb (2020) correlates below $0.11$ with every other source.
\end{minipage}
\end{table}

\clearpage

\begin{table}[h!]
\centering
\caption{Theorem 1 implementation}\label{tab:t2}
{\normalsize\itshape Panel A. Per-rater RESET curvature diagnostic}\\[6pt]
\small
\begin{tabular}{lcccccc}
\toprule
Rater & $\hat b_z$ & $\hat b_{z^2}$ & HC SE & HC $t$ & $\hat\gamma = 2\hat b_{z^2}$ & Reject linearity \\
\midrule
ChatGPT-5 & +0.951 & +0.060 & 0.022 & +2.73 & +0.120 & Yes \\
Claude 4.5 & +0.952 & -0.172 & 0.024 & -7.10 & -0.344 & Yes \\
GPT-4 (Eloundou et al. 2024) & +0.934 & -0.057 & 0.024 & -2.36 & -0.114 & Yes \\
Gemini 2.5 & +0.950 & +0.068 & 0.021 & +3.20 & +0.136 & Yes \\
\bottomrule
\end{tabular}
\\[6pt]
\begin{minipage}{0.95\textwidth}
\footnotesize\textbf{Notes.} Auxiliary RESET regression of each rater's standardized score on the leave-one-rater-out cross-source average and its square. The reported HC $t$-statistic uses heteroskedasticity-robust standard errors. All four raters reject linearity at the $1$ percent level, which confirms the nonlinear measurement regime under which the partial-identification framework operates. Twice the maximum absolute quadratic coefficient (Claude 4.5) equals $0.344$ in standardized units; this OLS-RESET bound is attenuated by errors-in-variables noise in the leave-one-out proxy, and the headline curvature bound used throughout the paper is the larger split-instrument IV-RESET estimate $\widehat B = 0.438$ of Panels B and C. Sample is $379$ six-digit SOC codes in the ACS analysis sample.
\end{minipage}
\end{table}

\clearpage

\begin{table}[h!]
\centering
\caption*{Table 2 (continued)}
{\normalsize\itshape Panel B. Theorem 1 empirical instantiation}\\[6pt]
\small
\begin{tabular}{lcc}
\toprule
Quantity & 4-LLM panel & 5-source panel \\
\midrule
$\max_r |\hat b_{z^2}^{(r)}|$ & 0.172 & 0.137 \\
$\widehat B$ (split-instrument IV-RESET, relative curvature) & 0.4380 & 0.4147 \\
$\widehat B^{\mathrm{OLS}} = 2 \max_r |\hat b_{z^2}^{(r)}|$ (OLS-RESET, attenuated) & 0.344 & 0.273 \\
$\hat\kappa_3$ (skewness, IV-RESET panel) & +0.1831 & +0.1331 \\
$\hat\kappa_4$ (kurtosis, IV-RESET panel) & 1.6486 & 1.5913 \\
$\hat\kappa_4 - 1 - \hat\kappa_3^2$ & 0.6151 & 0.5736 \\
$\widehat\Delta_{\mathrm{sym}} = \tfrac{1}{8}\widehat B^2(\hat\kappa_4 - 1 - \hat\kappa_3^2)$ & 0.0147 & 0.0123 \\
$\widehat\Delta_{\mathrm{sym}}$ as \% of point estimate & $1.47\%$ & $1.23\%$ \\
$\widehat\Delta_{\mathrm{sym}}$ at upper-95\% bound on $\widehat B$ & $2.47\%$ & $1.88\%$ \\
\bottomrule
\end{tabular}
\\[6pt]
\begin{minipage}{0.85\textwidth}
\footnotesize\textbf{Notes.} Symmetric Theorem 1 half-width $\widehat\Delta_{\mathrm{sym}}$ instantiated from the split-instrument IV-RESET relative-curvature bound (Section~4.2) and the sample third and fourth standardized moments of the cross-source average. The 4-LLM panel uses the four LLM raters under the Eloundou et al.\ (2024) rubric on the 379-SOC analysis sample. The 5-source panel augments with the Felten et al.\ (2021) ability-based score on the 682-SOC full-coverage sample. The partial-identification uncertainty ($1.47$ percent in the 4-LLM panel and $1.23$ percent in the 5-source panel, raised to $2.47$ and $1.88$ at the upper-95\% bound on $\widehat B$) is small relative to the typical sampling uncertainty on the $8.88$-million-worker ACS sample.
\end{minipage}
\end{table}

\begin{table}[h!]
\centering
\caption*{Table 2 (continued)}
{\normalsize\itshape Panel C. Per-rater split-instrument IV-RESET relative curvature}\\[6pt]
\small
\begin{tabular}{lcccc}
\toprule
 & \multicolumn{2}{c}{4-LLM panel} & \multicolumn{2}{c}{5-source panel} \\
\cmidrule(lr){2-3}\cmidrule(lr){4-5}
Rater & $\hat g_s$ & HC $t$ & $\hat g_s$ & HC $t$ \\
\midrule
ChatGPT-5 & $\mathbf{+0.438}$ & $+5.58$ & $\mathbf{+0.415}$ & $+6.97$ \\
Claude 4.5 & $-0.323$ & $-7.05$ & $-0.269$ & $-5.95$ \\
GPT-4 & $-0.215$ & $-5.34$ & $-0.088$ & $-2.07$ \\
Gemini 2.5 & $+0.094$ & $+2.08$ & $+0.248$ & $+5.38$ \\
Felten & --- & --- & $-0.236$ & $-4.37$ \\
\bottomrule
\end{tabular}
\\[6pt]
\begin{minipage}{0.85\textwidth}
\footnotesize\textbf{Notes.} Per-source split-instrument IV-RESET relative curvature $\hat g_s = 2 \hat b_2 \sqrt{\mathrm{Cov}(z_A, z_B)} / \hat b_1$ from the auxiliary regression of each standardized source on a leave-out subset average and its square, instrumented by a disjoint leave-out subset average and its square (Section~4.2, equation~15). HC $t$ uses heteroskedasticity-robust standard errors on the quadratic coefficient. Every source rejects linearity at the $1$ percent level. The curvature bound $\widehat B = \max_s |\hat g_s|$ is attained at ChatGPT-5 on both panels (boldface), in contrast to the OLS RESET of Panel A, whose maximum absolute quadratic coefficient is at Claude 4.5; the errors-in-variables correction reweights the raters.
\end{minipage}
\end{table}

\clearpage

\begin{landscape}
\begin{table}[h!]
\centering
\caption{Single-source vs MS-GMM corrected downstream estimates}\label{tab:t3}
\small
\begin{tabular}{llcccccc}
\toprule
Outcome & Estimator & $\hat\beta \times 100$ & SE & $t$ & 95\% CI (sampling) & Theorem 1 bands & Combined region \\
\midrule
\multirow{7}{*}{Employment}  & Single-source: ChatGPT-5 & $-0.197^{**}$ & $(0.084)$ & $-2.36$ & $[-0.361, -0.033]$ & --- & --- \\
                            & Single-source: Claude 4.5 & $-0.262^{***}$ & $(0.095)$ & $-2.75$ & $[-0.449, -0.075]$ & --- & --- \\
                            & Single-source: GPT-4 & $-0.258^{***}$ & $(0.087)$ & $-2.98$ & $[-0.428, -0.089]$ & --- & --- \\
                            & Single-source: Gemini 2.5 & $-0.206^{**}$ & $(0.082)$ & $-2.53$ & $[-0.366, -0.046]$ & --- & --- \\
                            & Single-source: Felten & $-0.223^{**}$ & $(0.090)$ & $-2.48$ & $[-0.400, -0.047]$ & --- & --- \\
                            & Single-source: Webb & $+0.023$ & $(0.092)$ & $+0.25$ & $[-0.157, +0.204]$ & --- & --- \\
                            & MS-GMM 5-source (symmetric) & $-0.239^{***}$ & $(0.090)$ & $-2.65$ & $[-0.417, -0.062]$ & $[-0.242, -0.236]$ & $[-0.419, -0.064]$ \\
\midrule
\multirow{7}{*}{LFP}  & Single-source: ChatGPT-5 & $+0.087$ & $(0.062)$ & $+1.41$ & $[-0.034, +0.208]$ & --- & --- \\
                            & Single-source: Claude 4.5 & $+0.081$ & $(0.073)$ & $+1.10$ & $[-0.063, +0.224]$ & --- & --- \\
                            & Single-source: GPT-4 & $+0.077$ & $(0.067)$ & $+1.16$ & $[-0.054, +0.209]$ & --- & --- \\
                            & Single-source: Gemini 2.5 & $+0.083$ & $(0.058)$ & $+1.43$ & $[-0.031, +0.196]$ & --- & --- \\
                            & Single-source: Felten & $+0.120^{*}$ & $(0.071)$ & $+1.69$ & $[-0.019, +0.259]$ & --- & --- \\
                            & Single-source: Webb & $+0.132^{*}$ & $(0.073)$ & $+1.81$ & $[-0.011, +0.276]$ & --- & --- \\
                            & MS-GMM 5-source (symmetric) & $+0.099$ & $(0.068)$ & $+1.45$ & $[-0.035, +0.232]$ & $[+0.098, +0.100]$ & $[-0.033, +0.234]$ \\
\midrule
\multirow{7}{*}{Log wage}  & Single-source: ChatGPT-5 & $-1.012^{*}$ & $(0.522)$ & $-1.94$ & $[-2.036, +0.012]$ & --- & --- \\
                            & Single-source: Claude 4.5 & $-1.178^{**}$ & $(0.541)$ & $-2.18$ & $[-2.238, -0.118]$ & --- & --- \\
                            & Single-source: GPT-4 & $-1.172^{**}$ & $(0.543)$ & $-2.16$ & $[-2.237, -0.107]$ & --- & --- \\
                            & Single-source: Gemini 2.5 & $-0.809$ & $(0.551)$ & $-1.47$ & $[-1.890, +0.271]$ & --- & --- \\
                            & Single-source: Felten & $-1.324^{***}$ & $(0.498)$ & $-2.66$ & $[-2.299, -0.349]$ & --- & --- \\
                            & Single-source: Webb & $-0.783$ & $(0.564)$ & $-1.39$ & $[-1.889, +0.323]$ & --- & --- \\
                            & MS-GMM 5-source (symmetric) & $-1.139^{**}$ & $(0.549)$ & $-2.08$ & $[-2.215, -0.064]$ & $[-1.153, -1.125]$ & $[-2.228, -0.076]$ \\
\bottomrule
\end{tabular}
\\[6pt]
\begin{minipage}{0.95\textwidth}
\footnotesize\textbf{Notes.} Difference-in-differences regression of the indicated labor-market outcome on the post-2022 interaction with the AI-exposure measure listed in the second column, estimated on the $8.88$-million-worker American Community Survey panel. The MS-GMM 5-source estimator is the symmetric cross-source half-average (minimum-distance) estimator of this paper. The combined region applies the Imbens--Manski construction with the Stoye (2009) critical value to the Theorem 1 endpoints, adding cluster-robust sampling margins with delta-method endpoint standard errors; it is defined only for the MS-GMM estimator, whose curvature band it extends. The factor-of-eleven spread across single-source employment elasticity estimates (from $+0.023$ on Webb to $-0.262$ on Claude 4.5) is reconciled by the MS-GMM consensus estimate at $-0.239$. Standard errors in parentheses are cluster-robust at the six-digit SOC level. Significance: $^{***}\,p<0.01$, $^{**}\,p<0.05$, $^{*}\,p<0.10$.
\end{minipage}
\end{table}
\end{landscape}

\clearpage

\begin{landscape}
\begin{table}[h!]
\centering
\caption{Convergence of multi-rater corrections (4-LLM panel)}\label{tab:t4}
\small
\begin{tabular}{llccc}
\toprule
Outcome & Estimator & $\hat\beta \times 100$ & SE & Partial-ID band \\
\midrule
\multirow{4}{*}{Employment}    & Dawid--Skene MLE & $-0.296$ & $(0.039)$ & --- \\
                                & ORIV stack-then-IV & $-0.303$ & $(0.084)$ & --- \\
                                & Imbens--Manski region & --- & --- & $[-0.394,\, -0.187]$ \\
                                & MS-GMM cross-source half-average 2SLS & $-0.264$ & $(0.088)$ & $[-0.268,\, -0.260]$ \\
\midrule
\multirow{4}{*}{LFP}    & Dawid--Skene MLE & $+0.034$ & $(0.034)$ & --- \\
                                & ORIV stack-then-IV & $+0.031$ & $(0.059)$ & --- \\
                                & Imbens--Manski region & --- & --- & $[-0.047,\, +0.092]$ \\
                                & MS-GMM cross-source half-average 2SLS & $+0.077$ & $(0.061)$ & $[+0.076,\, +0.078]$ \\
\midrule
\multirow{4}{*}{Log wage}    & Dawid--Skene MLE & $-1.451$ & $(0.105)$ & --- \\
                                & ORIV stack-then-IV & $-1.439$ & $(0.323)$ & --- \\
                                & Imbens--Manski region & --- & --- & $[-1.737,\, -0.916]$ \\
                                & MS-GMM cross-source half-average 2SLS & $-1.102$ & $(0.500)$ & $[-1.118,\, -1.086]$ \\
\bottomrule
\end{tabular}
\\[6pt]
\begin{minipage}{0.95\textwidth}
\footnotesize\textbf{Notes.} Each row reports the post-2022 difference-in-differences coefficient under one multi-rater correction framework on the four-language-model panel matched to the IPUMS-ACS analysis sample of $8{,}883{,}132$ person-years for $2015$ to $2024$, identical to the sample in Table~3. The Dawid--Skene MLE delivers the latent-class posterior membership probability that the downstream regression uses as the corrected exposure under conditional independence. The ORIV stack-then-IV estimator stacks the data four times (one per rater) and instruments each rater's score with the leave-one-rater-out average. The Imbens--Manski region is constructed from the four single-rater OLS estimates with the Stoye (2009) critical value $c = 1.65$, evaluated at the estimated standardized interval width $\hat r = 2.03$. The MS-GMM cross-source half-average estimator is the four-LLM variant, for direct comparability with the four-LLM discrete-limit corrections; the headline five-source consensus is in Table~3. The three discrete-limit point estimators agree with the MS-GMM smooth-curvature estimator in sign and order of magnitude on the employment and wage coefficients; on labor-force participation all four estimates are small and statistically indistinguishable from zero. The Imbens--Manski region contains the MS-GMM point estimate on every outcome, including labor-force participation, where the MS-GMM point ($+0.077$) lies inside the region ($[-0.047, +0.092]$). The Theorem 1 band reported for the MS-GMM estimator is a curvature interval, not a sampling confidence interval, and is not expected to contain the discrete-limit point estimates.
\end{minipage}
\end{table}
\end{landscape}

\clearpage
\section*{Figures}
\addcontentsline{toc}{section}{Figures}

\clearpage

\clearpage

\clearpage

\section*{Appendix Figures}
\addcontentsline{toc}{section}{Appendix Figures}
\renewcommand{\thefigure}{A\arabic{figure}}\setcounter{figure}{0}

\begin{figure}[h!]
\centering
\includegraphics[width=0.95\linewidth,keepaspectratio]{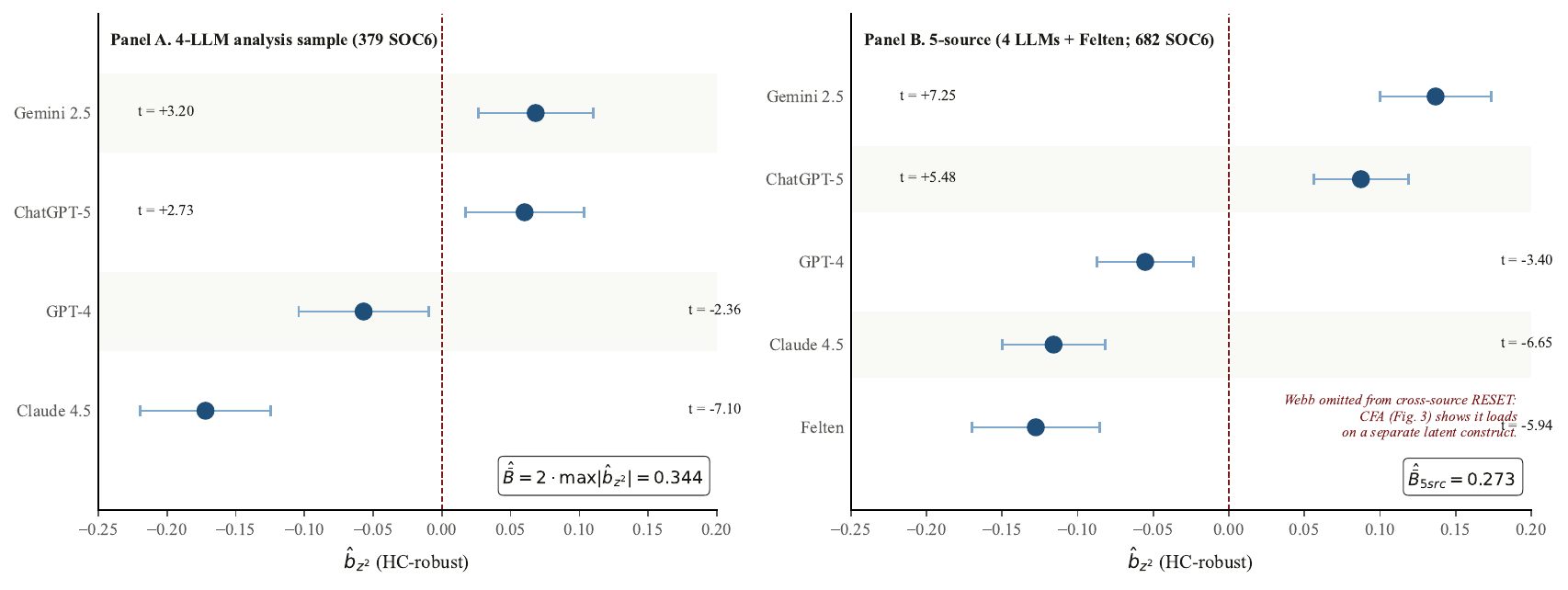}
\caption{Per-rater RESET curvature diagnostic, 4-LLM and 5-source panels.}\label{fig:a1}
\begin{minipage}{0.95\linewidth}
\footnotesize\textbf{Notes.} Quadratic coefficient $\hat b_{z^2}$ from the auxiliary regression $\tilde y^{(r)} = a + b_z z + b_{z^2} z^2 + e$, where $z$ is the leave-one-source-out cross-source average. Panel A reports the four LLMs on the 379-SOC6 analysis sample with the OLS-RESET bound $\hat{\bar B}^{\mathrm{OLS}} = 2 \cdot \max_r|\hat b_{z^2}| = 0.344$. Panel B reports the four LLMs and Felten on the 682-SOC6 panel with $\hat{\bar B}^{\mathrm{OLS}}_{5\mathrm{src}} = 0.273$. These OLS-RESET bounds are attenuated by errors-in-variables noise; the headline analysis uses the larger split-instrument IV-RESET bounds $\widehat B = 0.438$ and $0.4147$ respectively (Section 4.2). Webb is omitted from both panels because the confirmatory factor analysis of Figure 1B shows it loads on a separate construct.
\end{minipage}
\end{figure}

\clearpage

\begin{figure}[h!]
\centering
\includegraphics[width=0.95\linewidth,keepaspectratio]{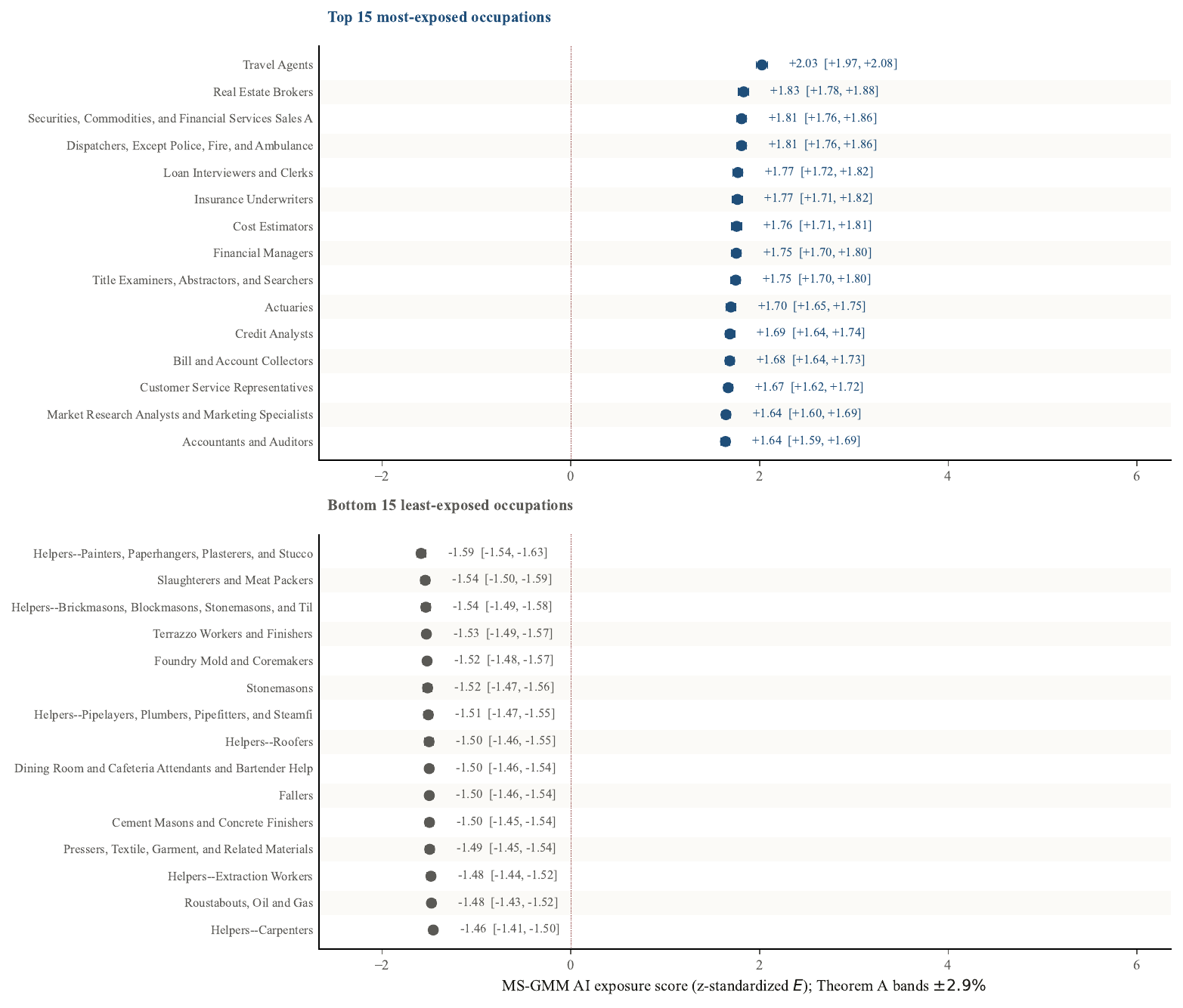}
\caption{Cross-source-corrected exposure score, top 15 and bottom 15 occupations.}\label{fig:a2}
\begin{minipage}{0.95\linewidth}
\footnotesize\textbf{Notes.} Cross-source-corrected MS-GMM exposure score $\hat E_{5\mathrm{src}}$ by six-digit Standard Occupational Classification code. The top 15 most-exposed occupations are professional and clerical categories with high LLM-rated applicability; the bottom 15 least-exposed are physical-labor categories. Error bars represent the Theorem 1 partial-identification interval $[\hat E(1-\hat\Delta), \hat E(1+\hat\Delta)]$ with the symmetric half-width $\hat\Delta = 0.0123$ in the 5-source panel.
\end{minipage}
\end{figure}

\clearpage

\begin{figure}[h!]
\centering
\includegraphics[width=0.95\linewidth,keepaspectratio]{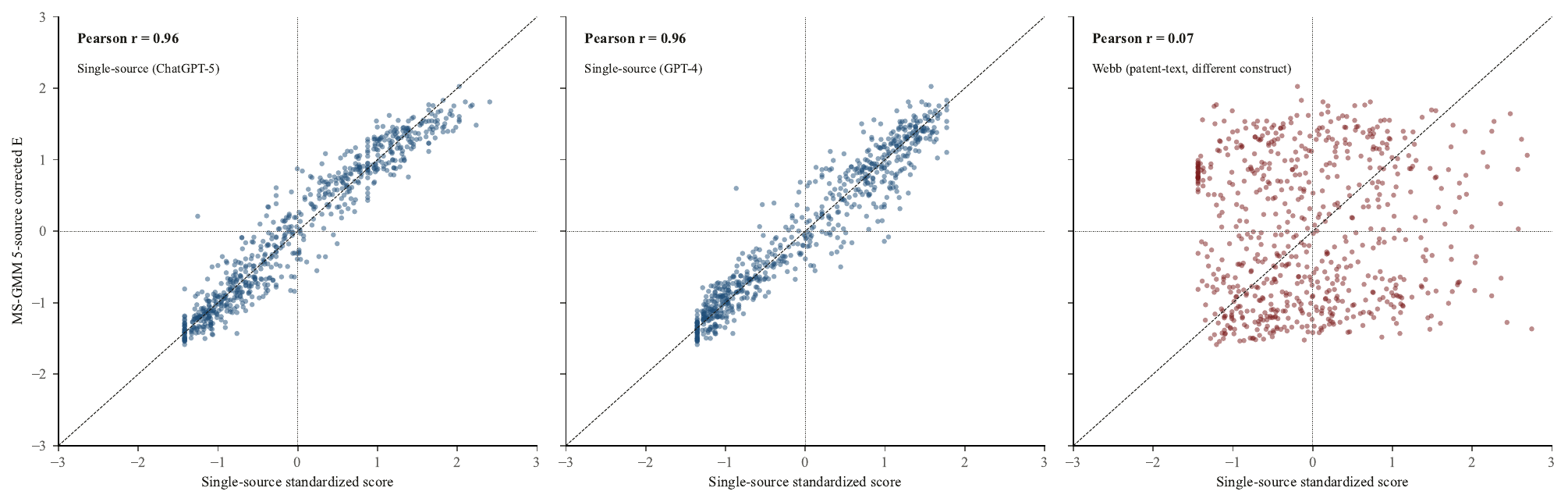}
\caption{Corrected MS-GMM score versus single-source standardized scores.}\label{fig:a3}
\begin{minipage}{0.95\linewidth}
\footnotesize\textbf{Notes.} Scatterplot of the cross-source-corrected MS-GMM score (vertical axis) against each single-source standardized score (horizontal axis) on the 682-occupation panel. The 45-degree dashed line indicates exact agreement. The LLM-rated single-source scores (ChatGPT-5, GPT-4) are highly correlated with the MS-GMM corrected score ($r = 0.96$), while the Webb patent-text score is essentially uncorrelated with it ($r = 0.07$), consistent with the confirmatory factor analysis result of Figure 1B.
\end{minipage}
\end{figure}

\clearpage

\begin{figure}[h!]
\centering
\includegraphics[width=0.95\linewidth,height=0.78\textheight,keepaspectratio]{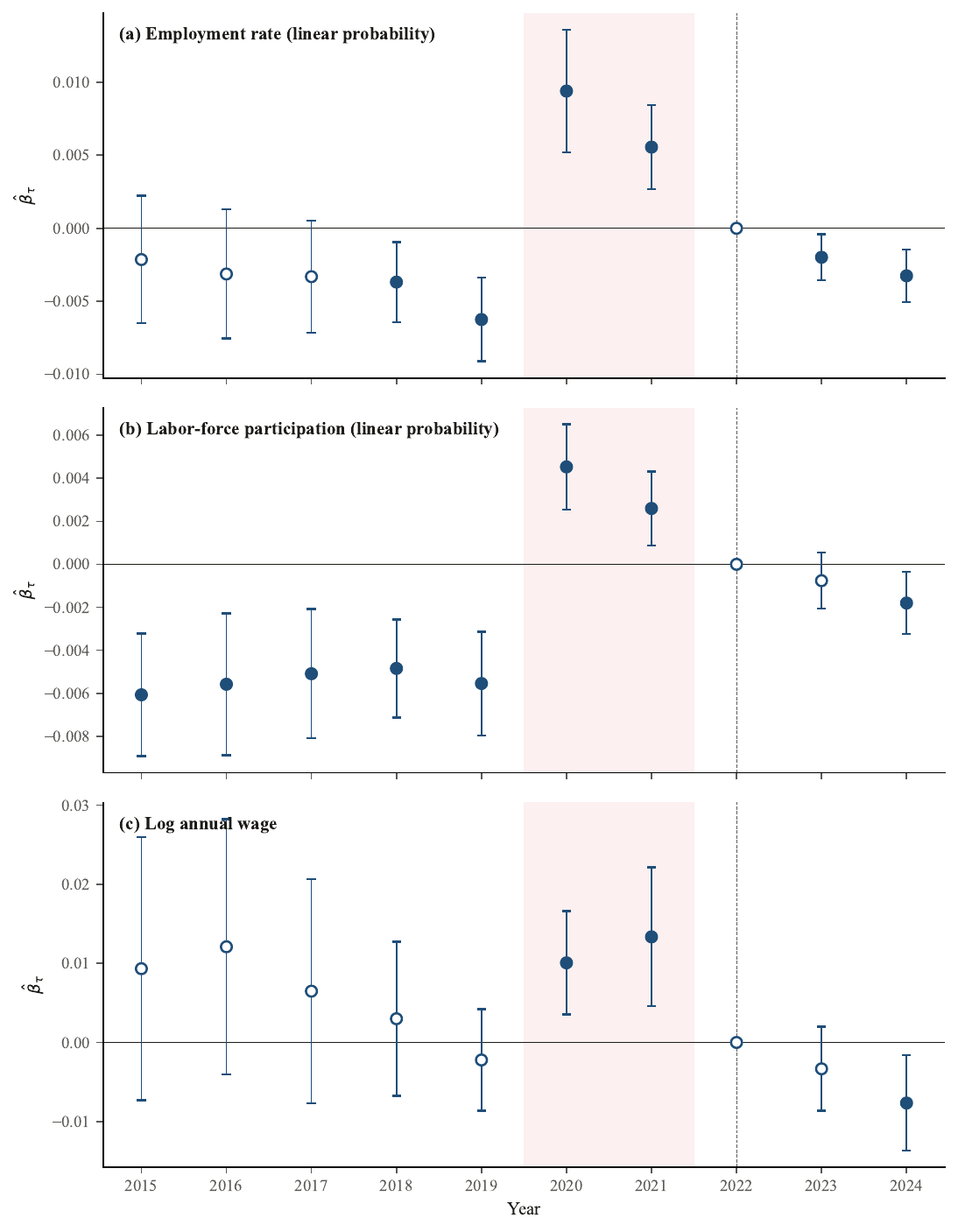}
\caption{Year-by-year event study with COVID-shaded period.}\label{fig:a4}
\begin{minipage}{0.95\linewidth}
\footnotesize\textbf{Notes.} Year-by-year coefficient $\hat\beta_\tau$ from the event-study regression $Y_{it} = \alpha_o + \delta_t + \sum_\tau \beta_\tau (E_o \times \mathbf{1}\{t = \tau\}) + \varepsilon_{it}$ with the reference year 2022. Panel (a) reports employment, panel (b) labor-force participation, and panel (c) log annual wage. Filled markers are coefficients significant at the 5\% level; hollow markers are not. The shaded region marks the COVID-pandemic years 2020-2021 in which the apparent post-2022 employment effect is partly driven by the remote-work composition shift.
\end{minipage}
\end{figure}

\clearpage

\begin{figure}[h!]
\centering
\includegraphics[width=0.95\linewidth,height=0.78\textheight,keepaspectratio]{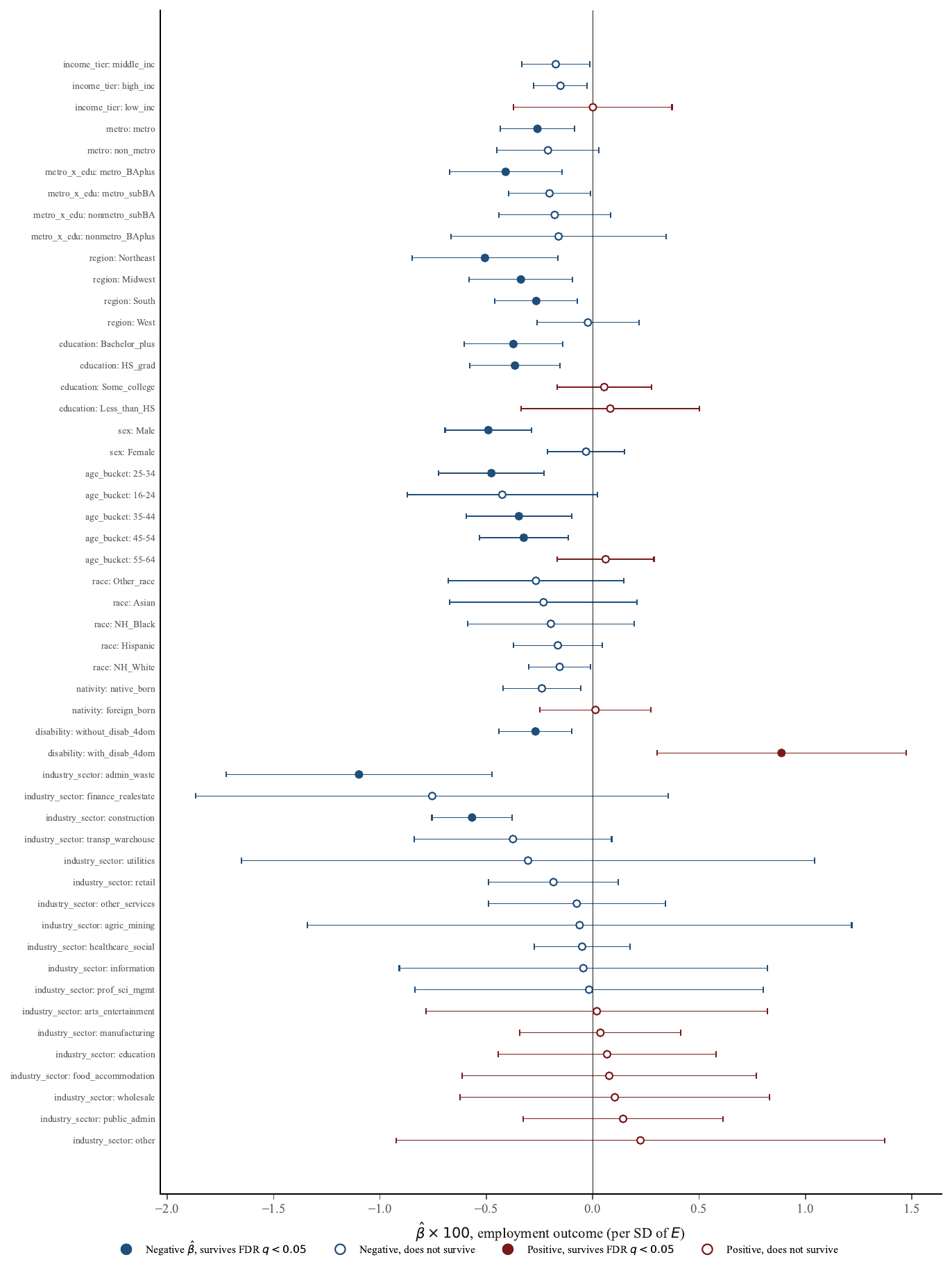}
\caption{Heterogeneity in post-2022 employment elasticity by subgroup.}\label{fig:a5}
\begin{minipage}{0.95\linewidth}
\footnotesize\textbf{Notes.} Forest plot of the post-2022 employment elasticity estimated separately by demographic and industry subgroups on the 8.88-million-worker ACS panel. Filled markers denote subgroups surviving Benjamini-Hochberg false-discovery-rate correction at $q < 0.05$ across the 153 subgroup-outcome combinations; hollow markers do not. The disability subgroup (4-domain definition) is the only positive employment-coefficient subgroup that survives the FDR correction.
\end{minipage}
\end{figure}

\clearpage

\begin{figure}[h!]
\centering
\includegraphics[width=0.95\linewidth,height=0.78\textheight,keepaspectratio]{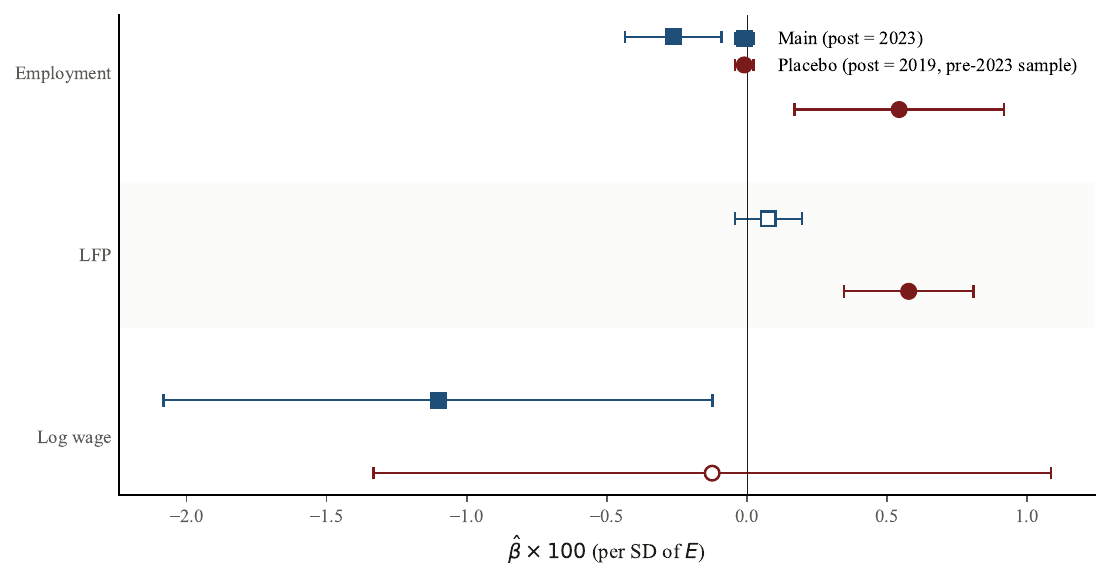}
\caption{Pre-treatment placebo comparison.}\label{fig:a6}
\begin{minipage}{0.95\linewidth}
\footnotesize\textbf{Notes.} Placebo specification with a fictitious post indicator set at year 2019 on the pre-2023 sample, compared to the main specification with the actual post-2022 indicator. The placebo coefficients on employment and labor-force participation are both positive and significant, which combined with the COVID shading in Figure A4 indicates that the apparent post-2022 effect reflects in part a continuation of the COVID-era reversal rather than a clean post-ChatGPT differential.
\end{minipage}
\end{figure}

\clearpage

\begin{figure}[h!]
\centering
\includegraphics[width=0.95\linewidth,keepaspectratio]{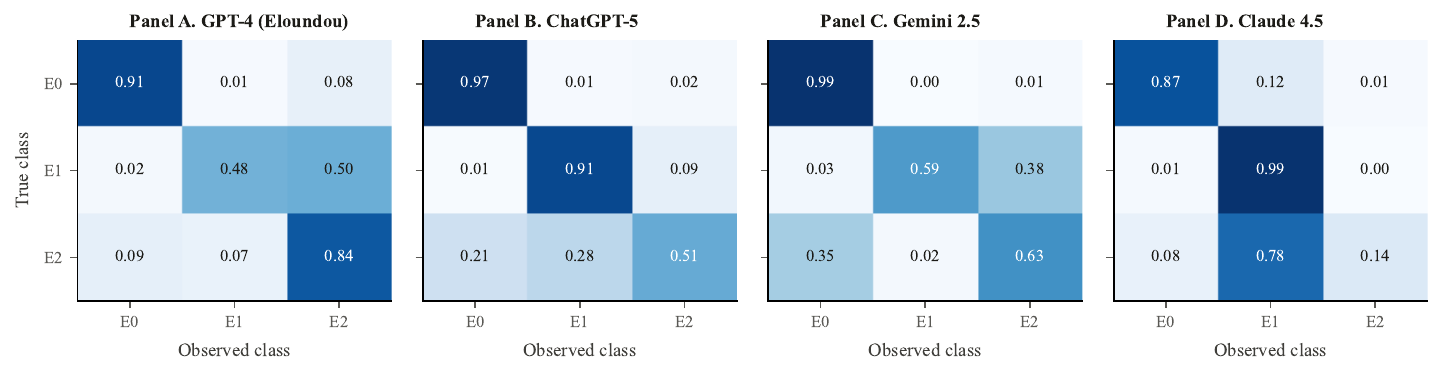}
\caption{Estimated $3 \times 3$ confusion matrices for the four LLM annotators.}\label{fig:a7}
\begin{minipage}{0.95\linewidth}
\footnotesize\textbf{Notes.} Each panel reports the rater-specific confusion matrix $C^{(k)}$ estimated by Dawid--Skene EM on the four-LLM annotation panel, with cell $(j, l)$ giving $P(Z_i^{(k)} = l \mid W_i = j)$ on a sequential blue scale. Diagonal entries report recall probabilities for each true class $E_0, E_1, E_2$ in the Eloundou et al.\ (2024) rubric. Off-diagonal mass quantifies the rater-specific misclassification structure that produces the rater attenuation factors $\hat\lambda^{(k)}$ discussed in Section 7.6 and that motivates the discrete-limit specialization of equation (2) introduced in Section 2. GPT-4 leads on $E_2$ recall (0.84). Claude 4.5 leads on $E_1$ recall (0.99) but classifies tool-mediated tasks ($E_2$) predominantly as $E_1$. Gemini 2.5 is the most conservative on $E_0$ (recall 0.99) but exhibits the highest probability of misclassifying true $E_2$ as $E_0$ (0.35), which produces its largest implied attenuation $\hat\lambda = 0.775$. ChatGPT-5 displays the most balanced profile across the three classes.
\end{minipage}
\end{figure}

\end{document}